\documentclass{aa}
\usepackage{graphicx}
\usepackage{hyperref}
\usepackage{txfonts}
\usepackage{physics}
\usepackage{siunitx}
\usepackage{float}
\usepackage[table]{xcolor}
\usepackage{amsmath}
\usepackage{orcidlink}
\usepackage{amsmath}
\usepackage[normalem]{ulem}

\newcommand{\lya}{$\mathrm{Ly}\alpha$}

\newcommand{\mhalo}{$M_\mathrm{h}$}
\newcommand{\vvir}{$v_\mathrm{vir}$}
\newcommand{\vmax}{$v_\mathrm{max}$}

\newcommand{\vmp}{$v_{M_\mathrm{peak}}$}
\newcommand{\thetab}{$\theta_\mathrm{beam}$}
\newcommand{\kms}{$\mathrm{km\ s^{-1}}$}

\begin{document}

\title{Three-dimensional stacking as a line intensity mapping statistic}
   \authorrunning{D.~A.~Dunne et al.}
   \titlerunning{Three-dimensional stacking as a line intensity mapping statistic}
   \author{
        D.~A.~Dunne\inst{1}\fnmsep\thanks{\email{\href{mailto:ddunne@astro.caltech.edu}{ddunne@astro.caltech.edu}}}\orcidlink{0000-0002-5223-8315}
        \and 
        K.~A.~Cleary\inst{1}\orcidlink{0000-0002-8214-8265}
        \and  
        P.~C.~Breysse\inst{2}\fnmsep\inst{3}\orcidlink{0000-0001-8382-5275}
        \and 
        D.~T.~Chung \inst{4}\orcidlink{0000-0003-2618-6504}
        \and 
        H.~T.~Ihle\inst{5}\orcidlink{0000-0003-3420-7766}
        \and 
        J.~G.~S.~Lunde\inst{5}\orcidlink{0000-0002-7091-8779}
        \and 
        H.~Padmanabhan\inst{6}\orcidlink{0000-0002-8800-5740}
        \and
        N.-O.~Stutzer\inst{5}\orcidlink{0000-0001-5301-1377}
        \and
        J.~R.~Bond\inst{7}\fnmsep\inst{8}\fnmsep\inst{9}\orcidlink{0000-0003-2358-9949}
        \and 
        J.~O.~Gundersen\inst{10}\orcidlink{0000-0002-7524-4355}
        \and 
        J.~Kim\inst{11}\orcidlink{0000-0002-4274-9373}
        \and
        A.~C.~S.~Readhead\inst{1}\orcidlink{0000-0001-9152-961X}
    }

   \institute{
       California Institute of Technology, 1200 E. California Blvd., Pasadena, CA 91125, USA
       \and 
       Center for Cosmology and Particle Physics, Department of Physics, New York University, 726 Broadway, New York, NY, 10003, U.S.A 
       \and 
       Department of Physics, Southern Methodist University, Dallas, TX 75275, USA
       \and 
       Department of Astronomy, Cornell University, Ithaca, NY 14853, USA
       \and 
       Institute of Theoretical Astrophysics, University of Oslo, P.O. Box 1029 Blindern, N-0315 Oslo, Norway
       \and 
       Departement de Physique Théorique, Universite de Genève, 24 Quai Ernest-Ansermet, CH-1211 Genève 4, Switzerland
       \and 
       Canadian Institute for Theoretical Astrophysics, University of Toronto, 60 St. George Street, Toronto, ON M5S 3H8, Canada
       \and 
       Department of Physics, University of Toronto, 60 St.~George Street, Toronto, ON, M5S 1A7, Canada
       \and 
       David A.~Dunlap Department of Astronomy, University of Toronto, 50 St.~George Street, Toronto, ON, M5S 3H4, Canada
       \and 
       Department of Physics, University of Miami, 1320 Campo Sano Avenue, Coral Gables, FL 33146, USA
       \and 
       Department of Physics, Korea Advanced Institute of Science and Technology (KAIST), 291 Daehak-ro, Yuseong-gu, Daejeon 34141, Republic of Korea
   }

   \date{Recieved 26 March 2025 / Accepted 8 August 2025}

    \abstract
    {
      Line intensity mapping (LIM) is a growing technique that measures the integrated spectral line emission from unresolved galaxies over a three-dimensional region of the Universe. Although LIM experiments ultimately aim to provide powerful cosmological constraints via auto-correlation, many LIM experiments are also designed to take advantage of overlapping galaxy surveys, thus enabling joint analyses of two datasets. We introduce a flexible simulation pipeline that can generate mock galaxy surveys and mock LIM data simultaneously for the same population of simulated galaxies. Using this pipeline, we explore a simple joint analysis technique: three-dimensional co-addition (stacking) of LIM data on the positions of galaxies from a traditional galaxy catalogue. We test how the output of this technique reacts to changes in experimental design of both the LIM experiment and the galaxy survey, its sensitivity to various astrophysical parameters, and its susceptibility to common systematic errors. We find that an ideal catalogue for a stacking analysis targets as many high-mass dark matter halos as possible. We also find that the signal in a LIM stacking analysis originates almost entirely from the large-scale clustering of halos around the catalogue objects rather than the catalogue objects themselves. While stacking is a sensitive and conceptually simple way to achieve a LIM detection, thus providing a valuable way to validate a LIM auto-correlation detection, it will likely require a full cross-correlation to achieve further characterisation of the galaxy tracers involved, as the cosmological and astrophysical parameters we explore here have degenerate effects on the stack.
    }


   \maketitle

\section{Introduction}\label{sec:intro}

Line intensity mapping (LIM) is an observational technique designed to measure the integrated spectral line emission from all galaxies in a population while generating a three-dimensional fluctuation map of a given spectral line with intentionally broadened spatial resolution. Specifically, this technique is sensitive to the integrated emission from even the faintest end of the luminosity function and is thus a possible solution to the completeness problem in galaxy surveys, particularly those at high redshift \citep[for a review, see][]{kovetz2019_limwhitepaper}. LIM is still an emerging field, and no experiment has yet made a confirmed auto-correlation detection of their target emission line \citep[e.g.][]{time2014_intro,keating2015_copss1,keating2020_mmime,cleary2021_comapoverview,fasano2022_concerto,ccatprime2023_overview,paul2023_meerkat} due to the faintness of the target signal, the need for exquisite control of instrumental systematic errors, and (in some cases) the presence of interloper spectral line emission.

Because of these difficulties, analysis techniques that combine LIM data with external tracers of the matter density (such as other intensity mapping experiments or more traditional galaxy surveys) will be an extremely valuable way of validating potential auto-correlation detections as the field matures. These joint analyses up-weight regions of the LIM map that are most likely to contain a signal and can therefore mitigate the effects of instrumental and astrophysical systematic errors, as these should primarily be unique to one of the two tracers included in the analysis. 

The most commonly studied technique for combining intensity mapping data with a galaxy survey is full three-dimensional cross-correlation \citep[e.g.][]{pullen2013_wmapquasars_limcrosscorr,chung2019_crosscorrelation,keenan2021_copssstack,breysse2019_agnfeedback}, as it allows for the study of the interaction between the two tracers as a function of spatial scale. However, this approach is sensitive to the existing angular structures in a survey strategy, and it suffers from complicated mode-mixing effects when the observational strategy of the LIM experiment is correlated with that of the galaxy catalogue. These problems are tractable but complex to manage, particularly in the early stages of an experiment. Because of this, several LIM experiments supplement their cross-correlation analyses with an analysis that is simpler conceptually, such as a three-dimensional stack (or co-addition) of the LIM data on the positions of objects in the galaxy survey \citep[e.g.][]{keenan2021_copssstack,dunne2024_ebossstacking,chen2025_meerklassstack}. Functionally, this is equivalent to the zero-lag region of the two-point correlation function, or a real-space version of a cross-correlation on the smallest spatial scale. Although this analysis does not probe the larger-scale modes that the cross-spectrum does, and thus loses some overall sensitivity compared to the cross-spectrum, it is easy to implement and is extremely useful for addressing systematic errors such as those listed above while still constraining the overall amplitude of the signal from a LIM tracer \citep{chen2025_meerklassstack}.

While the stack is conceptually simple, in practice it is generally unclear how astrophysical and experimental factors will affect its output. Previous work has shown that, for example, astrophysical line broadening has a dampening effect on cross-correlations that is worse in the higher-k bins, which are where the stack sensitivity lies \citep[e.g.][]{chung21_linebroadening}. These effects need to be explored in detail with simulations. Therefore, in this work, we devise a simple yet robust simulation pipeline that can generate a realistic population of multiple astrophysical tracers as well as parametrize the interaction between the two. Using this pipeline, we aim to answer the following questions:

\begin{enumerate}
    \item What experimental design (of both the LIM and catalogue experiment) is optimal for a stacking analysis?
    \item Where does the signal in a stacking analysis originate?
    \item What can the stack tell us about the properties of the galaxies contributing to the signal?
\end{enumerate}

We have structured this work as follows: In \S\ref{sec:simpipeline}, we introduce the joint simulation pipeline, which is an extension to the existing \verb|limlam_mocker|\footnote{\url{https://github.com/georgestein/limlam_mocker}} simulation pipeline known as \verb|joint_limlam_mocker.|\footnote{\label{note1}\url{https://github.com/delaneydunne/joint_limlam_mocker}} We briefly summarize the stacking methodology in \S\ref{sec:stacking_methods} and attempt to parametrize the stack analytically. We then test the effects of various instrumental and astrophysical parameters on the outcome of the stacking analysis in \S\ref{sec:simresults}. In \S\ref{sec:discussion}, we discuss the implications of these results. 

We assumed base ten logarithms unless otherwise stated, and where necessary, we took a $\Lambda$CDM cosmology with $\Omega_\mathrm{m} = 0.286$, $\Omega_\Lambda = 0.714$, $\Omega_\mathrm{b} = 0.047$, and $H_0=100h\mathrm{km\ s^{-1}\ Mpc^{-1}}$ with $h=0.7$. This matches the choice of parameters used in the peak-patch simulations in \S\ref{sec:simpipeline} and is based on the nine-year WMAP results \citep{Hinshaw_2013}. All cosmological distances are presented as comoving quantities.

\section{Simulated multi-tracer observations}\label{sec:simpipeline}

Because LIM targets cosmological scales but is also sensitive to variations in spectral line emission at the level of an individual dark matter (DM) halo, observational LIM data are difficult to simulate. For joint analyses such as the stack being explored in this work, we are additionally including a second spectral line tracer, and thus need to realistically simulate the distribution across halo masses of both tracers, as well as the interaction between the tracers at the individual halo level.

\begin{figure*}
    \centering
    \includegraphics[width=0.95\linewidth]{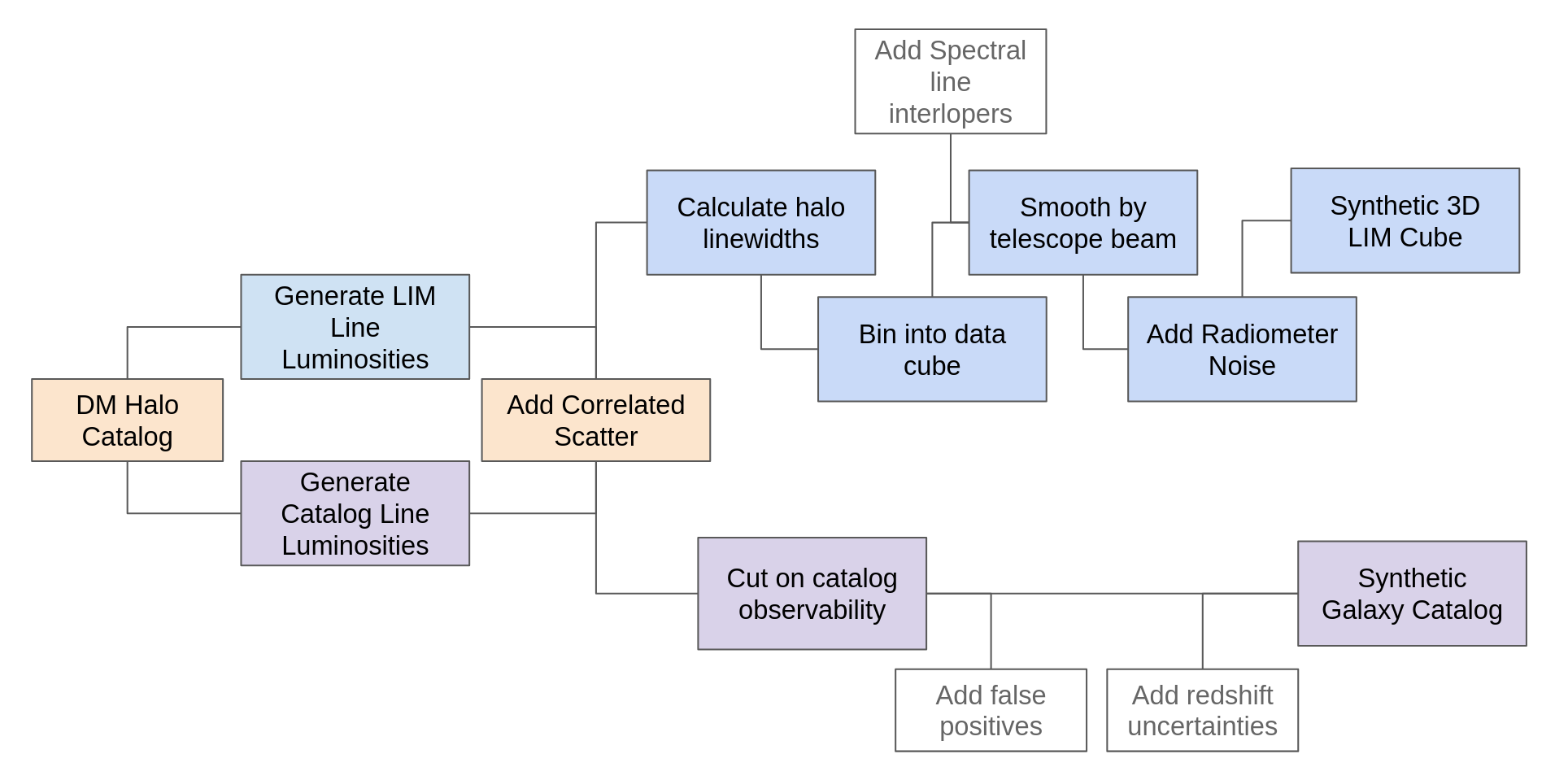}
    \caption{Flowchart depicting the multi-tracer simulation pipeline. Orange boxes indicate steps that affect both the galaxy catalogue and LIM data, blue boxes are actions on the simulated LIM data, and purple boxes are actions on the simulated galaxy catalogue. Steps with white boxes are optional.}
    \label{fig:simflowchart}
\end{figure*}

For this work we follow previous efforts to simulate LIM data, adding the capability to generate a galaxy catalogue using a different spectral line tracer. The resulting pipeline, known as \verb|joint_limlam_mocker|, is publicly available$^{\ref{note1}}$. The overall flow of the pipeline is outlined in Figure \ref{fig:simflowchart}, and is as follows: we begin with a DM halo catalogue generated from peak-patch $N$-body simulations (\S\ref{sec:simpipeline:peakpatchsims}). We then assign to each halo simulated luminosities corresponding to each of two different emission lines calculated based on the halo's DM mass -- one line associated with the LIM map (\S\ref{sec:simpipeline:co_luminosity_sim}) and one associated with the galaxy catalogue (\S\ref{sec:simpipeline:lya_luminosity_sim}). We add scatter to each luminosity value, correlated between the two tracers on a per-halo basis (\S\ref{sec:simpipeline:correlated_scatter}). We then generate a synthetic line intensity map (\S\ref{sec:simpipeline:simmap}) and a synthetic galaxy catalogue (\S\ref{sec:simpipeline:simgalaxycatalogue}). 

The outputs of the simulation pipeline are multi-survey and multi-tracer synthetic observations of the same population of simulated DM halos. The synthetic observations include a rudimentary treatment of the correlation between the tracers due to galaxy-scale astrophysics, as well as other complicating instrumental or astrophysical parameters, such as spectral line broadening, interloper emission in the LIM map, and redshift uncertainties in the galaxy catalogue. Although the treatment of these astrophysical and instrumental parameters is basic, we can vary the prescription used for any of the steps in this process, in order to explore the importance of each step to the stacking analysis. 

As an illustrative example, we based the instrumental and spectral line modelling parameters for the LIM experiment on the CO Mapping Array Project (COMAP) Pathfinder experiment \citep{cleary2021_comapoverview} and the parameters for the galaxy catalogue on the Hobby-Eberly Telescope Dark Energy eXperiment \citep[HETDEX,][]{gebhardt2021_hetdexoverview}, meaning that the two spectral lines we treat are the (1--0) transition of carbon monoxide (CO), and Hydrogen Lyman-$\alpha$ (Ly$\alpha$) respectively. We thus refer to the simulated luminosity values generated for each halo as the `CO luminosity' ($L_\mathrm{CO}$), which is used to generate the mock LIM data cube, and the `Ly$\alpha$ Luminosity' ($L_\mathrm{Ly\alpha}$), which is used to generate the resolved mock galaxy catalogue. We note, however, that this framework should be generic to any joint LIM-galaxy analysis. We explain the choices of the various parameters chosen for the default simulation setup below and summarize them in Table \ref{tab:LIM_params}.

\subsection{Peak-patch $N$-body simulations}\label{sec:simpipeline:peakpatchsims}
\begin{figure*}
    \centering
    \includegraphics[width=0.98\linewidth]{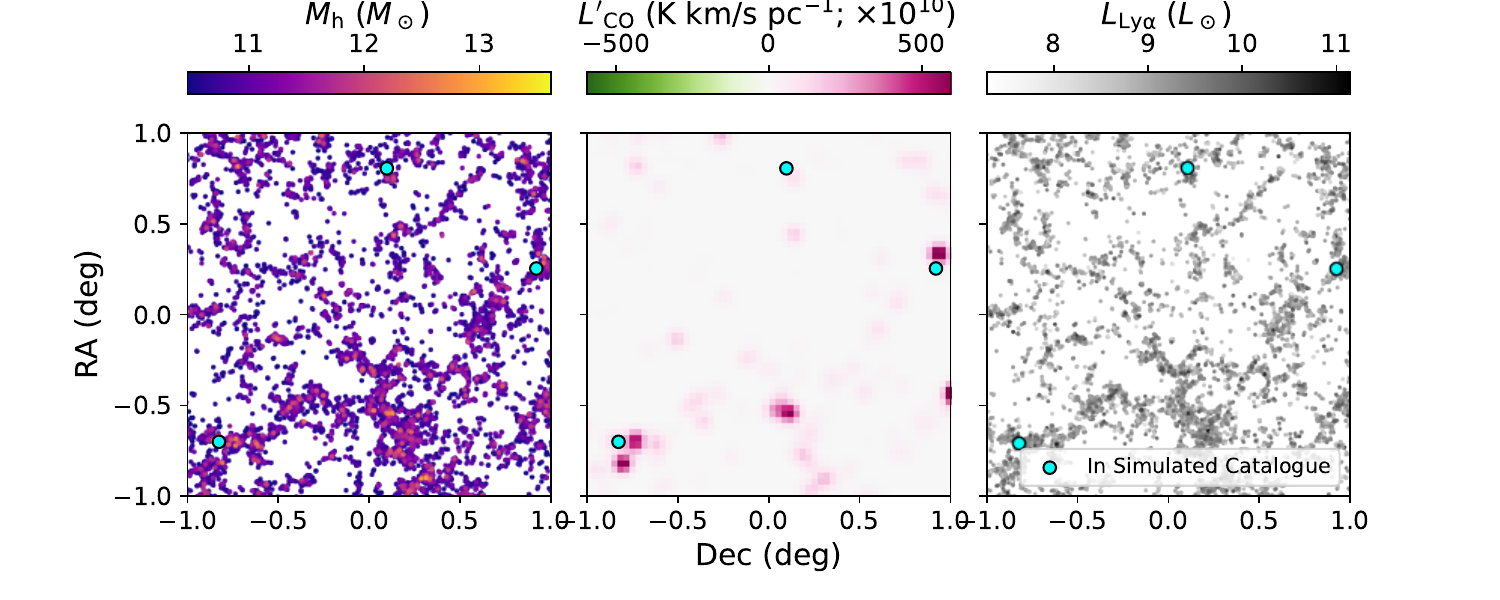}
    \caption{Zoomed-in frequency slices of the simulated population of DM halos and resulting mock observations. \textit{Left:}DM halos, coloured by their halo mass. \textit{Centre:} Mock LIM fluctuation map of the CO emission (with no noise added). \textit{Right:} Ly$\alpha$ luminosity of each DM halo. The halos that would actually be detected by the mock survey and thus included in the galaxy catalogue are shown as larger cyan circles in all three panels.}
    \label{fig:joint_sim_slices}
\end{figure*}

The basis of the synthetic observations is a set of $N$-body simulations generated using the peak-patch method \citep{bondmeyers1996_peakpatchsims, stein2019_peakpatchsims}. These simulations provide a realistic catalogue of the three-dimensional (3D) positions and masses of DM halos, onto which luminosities can be painted. The minimum halo mass we include is $2.9\times10^{10}\ M_\odot$, and the maximum mass is $9.1\times10^{13}\ M_\odot$, following \citet{chung2021_comapforecasts}. A cutout of the resulting distribution of DM halos is shown in Figure \ref{fig:joint_sim_slices}.

\subsection{LIM tracer luminosity modelling: CO(1--0)}\label{sec:simpipeline:co_luminosity_sim}
The prescriptions we used to determine the CO luminosity of the simulated halos are the same as those used for our COMAP auto-correlation analyses, to enable direct comparison. Each of these models calculates the CO luminosity of a DM halo directly from its mass. They are:

\begin{itemize}
    \item The data-driven `COMAP fiducial' model \cite[][hereafter C22]{chung2021_comapforecasts}, which we treated as the default for these simulations. It follows the functional form 
    \begin{equation}
        \frac{L'_\mathrm{CO}(M_\mathrm{h})}{\mathrm{K\ km\ s^{-1}\ pc^2}} = \frac{C}{(M_\mathrm{h}/M)^A + (M_\mathrm{h}/M)^B},
    \end{equation}
    where $A$, $B$, $C$, and $M$ are all free parameters. \citet{chung2021_comapforecasts} determine fiducial values for these free parameters based on the \textsc{UniverseMachine} (UM) framework of \citet{behroozi2019_universemachine} and conditioned on data from COLDz \citep{riechers2019_coldz} and COPSS \citep{keating2016_copss2}, which are listed in Table \ref{tab:LIM_params}.
    \item The CO model from \cite{padmanabhan2018_comodel}, which was abundance-matched onto CO observations at $z=3$ from \cite{keating2016_copss2}. The model takes the form of a double power law with redshift-dependent parameters: 
    \begin{equation}
        L_\mathrm{CO}(M_\mathrm{h},z) = \frac{2N(z)M_\mathrm{h}}{\left(\frac{M_\mathrm{h}}{M_1(z)} \right)^{-b(z)} + \left(\frac{M_\mathrm{h}}{M_1(z)} \right)^{y(z)}}.
    \end{equation}
    Here, the parameters $M_1(z)$, $N(z)$, $b(z)$, and $y(z)$ all contain a constant term for $z\sim 0$ and a term that evolves with redshift:
    \begin{eqnarray}
        \log M_1(z) &=& \log M_{10} + M_{11}z/(z+1) \notag \\
        N(z) &=& N_{10} + N_{11} z/(z+1) \notag \\
        b(z) &=& b_{10} + b_{11}z/(z+1) \notag \\
        y(z) &=& y_{10} + y_{11} z/z(z+1).
    \end{eqnarray}
    We took the best-fitting parameters from Table 1 of \cite{padmanabhan2018_comodel}. This model includes an additional factor quantifying the duty cycle of DM halos (the percentage of which are actually emitting in CO), $f_\mathrm{duty}$. Here, we took $f_\mathrm{duty}=0.1$. We refer to this model hereafter as P18. 
    \item The CO model from \cite{li2016_comodelling}, which determines the CO luminosity of a given DM halo via its infrared luminosity calculated through a modelled star formation rate (SFR),
    \begin{equation}
        \log L'_\mathrm{CO} = \frac{1}{\alpha}\left[ \log L_\mathrm{IR} - \beta \right]
    \end{equation}
    where
    \begin{equation}
        L_\mathrm{IR} = \frac{\mathrm{SFR}}{\delta_\mathrm{MF}}\times 10^{10}.
    \end{equation}
    The coefficient $\delta_\mathrm{MF}$ depends on the initial mass function, and we calculated SFR values following \cite{behroozi2013_sfr1, behroozi2013_sfr2}. As in \cite{li2016_comodelling}, we took $\delta_\mathrm{MF} = 1$. For CO(1--0), 
    \begin{equation}
        \frac{L_\mathrm{CO}}{L_\odot} = 4.9\times 10^{-5} \frac{L'_\mathrm{CO}}{\mathrm{K\ km\ s^{-1}\ pc^2}}.
    \end{equation}
    We refer to this model as L16.
\end{itemize}

\begin{figure}
    \centering
    \includegraphics[width=0.95\linewidth]{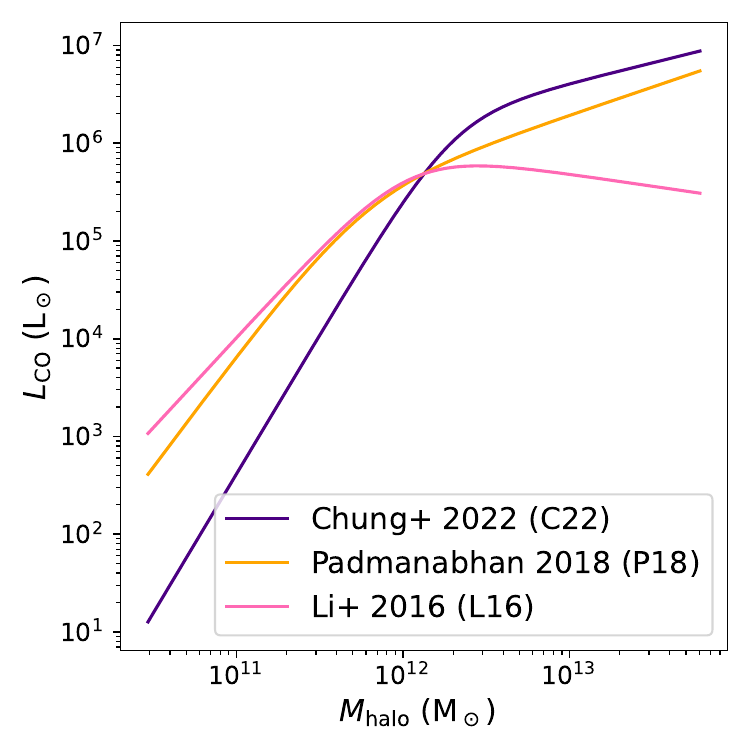}
    \caption{CO luminosity as a function of halo DM mass for the different CO models tested. The models are listed in \S\ref{sec:simpipeline:co_luminosity_sim}.}
    \label{fig:co_lum_funcs}
\end{figure}

For each of these different models, we show the CO luminosity $L(M_\mathrm{h})$ as a function of halo mass $M_\mathrm{h}$, in Figure \ref{fig:co_lum_funcs}. We note that there is more than an order of magnitude variation between the models at most halo masses -- this is a reflection of the large uncertainty that exists in this model space.

\subsection{Catalogue tracer luminosity modelling: Ly$\alpha$}\label{sec:simpipeline:lya_luminosity_sim}
We used luminosity functions to determine the Ly$\alpha$ luminosity of the DM halos, as these tend to be better studied than models directly connecting luminosity of galaxies to their host DM halo masses (i.e. $L(M_\mathrm{h})$ models) for optical line tracers such as Ly$\alpha$. We abundance-matched these luminosity functions onto the simulated DM mass function in order to assign to each DM halo a catalogue luminosity. For the purposes of this work, we assumed Schechter luminosity functions \citep{schechter1976}, 
\begin{equation}
    \frac{dn}{dL} = \frac{\phi^*}{L^*} \left(\frac{L}{L^*}\right)^\alpha e^{-L/L^*},
\end{equation}
where $L^*$ is a characteristic luminosity, $\phi^*$ is the normalisation density, and $\alpha$ is the faint-end power law slope. We varied only the input parameters here, although any functional form could be used for the catalogue luminosity function in the \verb|joint_limlam_mocker| framework.

\begin{figure}
    \centering
    \includegraphics[width=0.95\linewidth]{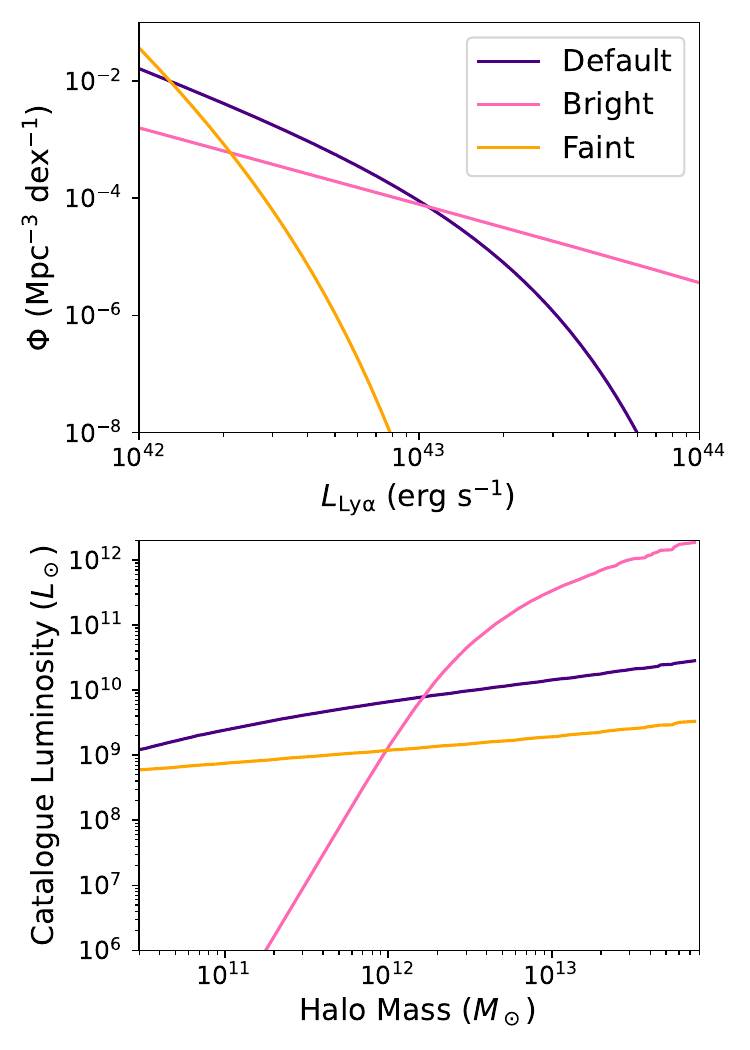}
    \caption{\textit{Top:} Various luminosity functions used to model the $L_\mathrm{Ly\alpha}$ assigned to each DM halo to generate the galaxy catalogue being stacked. Each model is a Schechter function, with parameters described in \S\ref{sec:simpipeline:lya_luminosity_sim}. \textit{Bottom:} Resulting luminosity as a function of DM halo mass.}
    \label{fig:catalogue_lum_funcs}
\end{figure}

We tested three different models for the Ly$\alpha$ luminosity function:
\begin{enumerate}
    \item The Lyman-$\alpha$ Emitter (LAE) luminosity function at $z=3.1$ from \cite{ouchi2020ARA&A_lya_lumfunc_review} based on observations from \cite{ouchi2008ApJS..176..301O} and \cite{konno2016bright}. These should be a fairly accurate estimate of the HETDEX luminosity function, as HETDEX probes LAEs in the same redshift range. This is the model we treated as the default. 

    \item The quasar Ly$\alpha$ luminosity function calculated at $z \sim 3$ using the Physics of the Accelerating Universe Survey \citep[PAUS; ][]{torralba-torregrosa2024_quasarlyalumfunc}. This model is much more bright-end heavy than typical LAE luminosity functions at this redshift range. Several large quasar catalogues (such as eBOSS, \citealt{eboss_dr16} or DESI, \citealt{chaussidon2023_DESIquasars}) may make interesting potential targets for stacking, so this is an interesting regime to probe. These catalogues use different spectral lines but trace similar populations of DM halos. We refer to this as our `bright' model throughout.

    \item A lower-redshift ($z = 0.3$) Ly$\alpha$ luminosity function from \cite{ouchi2008ApJS..176..301O}. This is a luminosity function for a galaxy population dominated by many fainter galaxies, and we used it as an example case for these types of populations. We refer to it as the `faint' model. 
\end{enumerate}
The Schechter parameters for each model are listed in Table \ref{tab:schechter_params}, and the luminosity functions themselves are compared in Figure \ref{fig:catalogue_lum_funcs}. For each choice of Ly$\alpha$ function, we used a scatter $\sigma_\mathrm{Ly\alpha} = 0.41$ dex, determined from the uncertainty in the Schechter parameters of the default model.

\begin{table}[]
    \caption{\label{tab:schechter_params} Schechter parameters for the catalogue luminosity function models.}
    \centering
    \begin{tabular}{ccccc}
    \hline
       Model  & $L^*$ & $\phi^*$ & $\alpha$ \\
            & (erg s$^{-1}$) & (Mpc$^{-3}$) & \\
       \hline
       Default & $8.49\times 10^{42}$ & $3.9\times10^{-4}$ & -1.8 \\
       Bright & $9.77\times10^{44}$ & $1.9\times10^{-7}$ & -1.34 \\
       Faint & $7.10\times10^{42}$ & $1.12\times10^{-4}$ & -1.8 \\
       \hline
    \end{tabular}
    \tablefoot{`Default' and `faint' values are taken from \cite{ouchi2020ARA&A_lya_lumfunc_review}, and `bright' values from \cite{torralba-torregrosa2024_quasarlyalumfunc}.}
\end{table}

\subsection{Correlated scatter in luminosities}\label{sec:simpipeline:correlated_scatter}
While the above formalism simulates how the two tracer luminosities vary with halo properties such as DM mass (thus roughly recreating the astrophysical bias for each tracer), it does not account for relationships between the two tracers in individual galaxies, which will set the level of the (cross-)shot noise. To account for this effect, we calculated the final luminosity values for each halo by introducing a correlated log-normal scatter $d$ between the luminosity determined from the \mhalo\ relations for the LIM tracer (here $L_\mathrm{CO}$) and the tracer of the resolved galaxy catalogue (here $L_\mathrm{Ly\alpha}$). This was done by scaling the luminosity values output from the previous steps by an exponential factor, 
\begin{equation}\label{eqn:scattered_lums}
    \left(\begin{array}{c}
        L_\mathrm{CO}\\
        L_\mathrm{Ly\alpha}
    \end{array} \right) =
    \left(\begin{array}{c}
        e^{\left[d_\mathrm{CO} - \sigma_\mathrm{CO}^2/2 \right]} L_\mathrm{CO} (M_\mathrm{h}) \\
        e^{\left[d_\mathrm{Ly\alpha} - \sigma_\mathrm{Ly\alpha}^2/2 \right]} L_\mathrm{Ly\alpha} (M_\mathrm{h})
    \end{array}\right),
\end{equation}
where $\sigma_\mathrm{CO}$ and $\sigma_\mathrm{Ly\alpha}$ are the base-$e$ standard deviations calculated from base ten values in decimal exponents (dex) as
\begin{equation}
    \sigma_\mathrm{CO} = \sigma_\mathrm{CO,\ dex}\ln 10,\ \sigma_\mathrm{Ly\alpha} = \sigma_\mathrm{Ly\alpha,\ dex} \ln 10
\end{equation}
($\sigma_\mathrm{CO,\ dex}$ and $\sigma_\mathrm{Ly\alpha,\ dex}$ are the amounts of logarithmic scatter in the CO and Ly$\alpha$ values, respectively, and are choices of the models used). The scaling factors $d_\mathrm{CO}$ and $d_\mathrm{Ly\alpha}$ were pulled from a two-dimensional normal distribution with zero mean and covariance given by 
\begin{equation}\label{eqn:correlated_scatter}
    \begin{split}
        \left(\begin{array}{c}
            d_\mathrm{CO}\\
            d_\mathrm{Ly\alpha}
        \end{array} \right) \leftarrow & \mathcal{N}\left\{ 
        \left(\begin{array}{c}
            0 \\
            0
        \end{array}\right),
        \left(\begin{array}{cc}
            \sigma^2_\mathrm{CO} & \rho \sigma_\mathrm{CO} \sigma_\mathrm{Ly\alpha} \\
            \rho \sigma_\mathrm{CO} \sigma_\mathrm{Ly\alpha} & \sigma^2_\mathrm{Ly\alpha}
        \end{array} \right)\right\}.
    \end{split}
\end{equation}
The parameter $\rho$ (where $-1<\rho < 1$) is the correlation coefficient, and it scales how the scatter in the two tracers relates for a given halo. 

For the example case we are exploring (a stack on a CO LIM map using a Ly$\alpha$ catalogue) an empirically motivated choice of $\rho$ is difficult due to the lack of available simultaneous observations of CO and Ly$\alpha$ in galaxies (especially at higher redshifts). For the purposes of this work, we looked at the relationship between CO and Ly$\alpha$ via galaxy metallicity. \cite{davis2023_HETDEXstack} found the LAEs catalogued by HETDEX to be primarily composed of galaxies with young metal-poor stellar populations ($\mathrm{Z} \sim 0.2-0.3\ \mathrm{Z_\odot}$). This agrees with other surveys of (especially lower-luminosity) LAEs \citep[e.g.][]{sobral2018_laes, matthee2021_xshooterlaes}. CO emission, however, is typically suppressed in actively star-forming low-metallicity galaxies \citep{bolatto13_cotoh2}. We can thus expect at least a slight anti-correlation between emission in the two lines on an individual galaxy level, no matter how their overall populations correlate with DM halo mass. With this as justification, we defaulted to a value of $\rho = -0.5$ for the purposes of this paper. We note that, particularly for our default CO and Ly$\alpha$ configuration, this choice does not significantly affect our results -- we explore the effects of varying $\rho$ in \S\ref{sec:simresults:astrophysics:rho}.

\subsection{Simulated intensity map}\label{sec:simpipeline:simmap}

We then generated a synthetic line intensity map, basing the simulated experimental parameters for this map on COMAP. We first simulated astrophysical sources of spectral line broadening by calculating a velocity to be associated with each halo. We used one of three methods:
\begin{enumerate}
    \item The relationship between peak halo mass (essentially identical to the halo mass at these redshifts) and \vmp\ provided by \cite{behroozi2019_universemachine} in their 
    UM framework,
    \begin{equation}
        v_{M_\mathrm{peak}}(M_\mathrm{h}) = (200\mathrm{\ km\ s^{-1}})\left[\frac{M_\mathrm{h}}{M_\mathrm{200\ km\ s^{-1}}(a)} \right]^{0.3},
    \end{equation}
    where $a$ is the scale factor at redshift $z$ and 
    \begin{equation}
        M_\mathrm{200\ km\ s^{-1}} = \frac{1.64\times10^{12} M_\odot}{(a/0.378)^{-0.142} + (a/0.378)^{-1.79}}.
    \end{equation}
    We added an additional log-normal scatter of $0.1$ dex, and we scaled the velocity of each halo with a randomly generated galaxy inclination, $i$:
    \begin{equation}
        v_\mathrm{max} = v_{M_{peak}}\sin i / 0.866.
    \end{equation}
    This is the prescription that aligns most closely with observations (see Appendix \ref{app:line_broadening}), and thus the one we treated as the default.
    
    \item The virial velocity (\vvir) associated with the DM mass of each halo:
    \begin{equation}
    \begin{split}
        v_\mathrm{vir} &= \sqrt{\frac{GM_\mathrm{h}}{r_\mathrm{vir}}} \\
          &= \left(\frac{\Delta_c}{2}\right)^{1/6} \left[G M_\mathrm{h} H(z) \right]^{1/3} \\
          &\approx 35\ \mathrm{km\ s^{-1}} \left(\frac{\Delta_c}{2}\right)^{1/6} \\ & \times \left(\frac{M_\mathrm{h}}{10^{10}\ M_\odot}\frac{H(z)}{100\ \mathrm{km\ s^{-1}\ Mpc^{-1}}}\right)^{1/3}.
    \end{split}
    \end{equation}
    This was again scaled by a randomly generated galaxy inclination $\sin i$.

    \item The virial velocity, calculated as above, but with a flat cut-off implemented at $v_\mathrm{max} = 1000 \mathrm{km\ s^{-1}}$. This primarily affects the largest DM halos ($M_\mathrm{h} \gtrsim 10^{12}\ M_\odot$), which can have virial velocities much larger than those seen in CO linewidths in observations (see Appendix \ref{app:line_broadening}, or \citealt{carilliwalter2013_highzmoleculargas}). Because the halos included in the stack are filtered through their catalogue luminosity, which correlates positively with mass for all catalogue models, these are the halos most likely to be included in the stack, and thus this cut-off may make a significant difference in the stacked linewidth.
\end{enumerate}

We compare these different strategies for calculating halo linewidths, as well as our choice of \vmp\ as a default prescription, in Appendix \ref{app:line_broadening}. The effects of each on the stack output are shown in \S\ref{sec:simresults:astrophysics:linebroadening}.

Following \citet{chung21_linebroadening}, we binned halos into $N_\mathrm{bin}$ linearly spaced bins by their \vmax. We then generated a CO intensity cube separately for each velocity bin, summing the halo luminosities into voxels (three-dimensional pixels) of sizes $\delta_x \times \delta_x \times \delta_\nu = 2\ \mathrm{arcmin}\ \times\ 2\ \mathrm{arcmin}\ \times\ 31.25\ \mathrm{MHz}$ ($\sim 3.7 \times 3.7 \times 4.1\ \mathrm{Mpc}$, or $k_\perp \sim 1.69\ \mathrm{Mpc^{-1}}$ and $k_\parallel \sim 1.53\ \mathrm{Mpc^{-1}}$). The intensity cubes cover $\Delta_x \times \Delta_x = 4^\circ \times 4^\circ$ total in their angular coordinates, and $\Delta_\nu = 8\ \mathrm{GHz}$ spectrally. We smoothed the intensity cube corresponding to each velocity bin along the line of sight by a Gaussian kernel with width given by the median \vmax\ in the bin, and summed together each smoothed cube. We approximated beam smoothing by convolving each spectral channel with a 2D Gaussian kernel with a full width at half maximum (FWHM) of \thetab\, defaulting to $\theta_\mathrm{beam} = 4.5$ arcmin ($\sim 8.3\ \mathrm{Mpc}$, or $k_\perp \sim 0.75\ \mathrm{Mpc^{-1}}$). In order to differentiate the LIM signal from the CO luminosity of a given halo (as the LIM signal comes from many halos summed together), we refer to the luminosity in the resulting map as $L_\mathrm{LIM}$. We show a frequency slice of a mock LIM cube at this stage (i.e.~noiseless) in Figure \ref{fig:joint_sim_slices}.

We also inserted varying amounts of simulated white noise into the map. We based the noise model on radiometer noise for the COMAP experiment, which is a focal-plane array with 19 feeds. Thus, assuming purely Gaussian noise (and even coverage over the simulated field), the COMAP noise response in each voxel is
\begin{equation}\label{eqn:comap_radiometer}
    \sigma_\mathrm{vox} = \frac{T_\mathrm{sys}}{\sqrt{2 \delta_\nu N_\mathrm{f} \tau/ \left(\Delta_x/ \delta_x\right)^2 }},
\end{equation}
where $T_\mathrm{sys}$ is the system temperature of the instrument, $\delta_\nu$ is the frequency width of each spectral channel, $N_\mathrm{f}$ is the number of feeds, $\tau$ is the total integration time of the experiment, and $\left(\Delta_x/ \delta_x\right)^2$ is the number of spaxels (spatial pixels) that coverage of the field is split across, assuming a scanning strategy with uniform coverage across the map. The values we used for each parameter are listed in Table \ref{tab:LIM_params}, and were based on the COMAP Pathfinder telescope \citep[e.g.][]{cleary2021_comapoverview}. However, to ensure we have high-significance stacks to analyse, we used a predicted per-field integration time $\tau$ that corresponds to a proposed future evolution of COMAP in which the currently operating Pathfinder instrument and two duplicates each observe for a total of ten years. This results in 29 000 hours spent integrating on each of the three COMAP fields \citep[c.f.][]{breysse2021_comapEOR}. We added a random value pulled from a distribution with this standard deviation and a zero mean to each voxel.

Additionally, we simulated line fluctuations from foreground or background populations of galaxies, which may be redshifted to the same observed frequency as the LIM data. For COMAP, this interloping emission will likely be CO(2-1) from galaxies at $6<z<8$. For [CII]-based surveys such as CONCERTO, TIME, or FYST, interloper emission will come from other $\sim1\ \mathrm{mm}$ emission lines, including [CI] and the CO rotational ladder \citep[e.g.][]{bethermin2022_concertomodeling}. To mimic this emission, we added simulated interloping spectral line emission to the map. For efficiency, we generated the interloper map by applying one of several linear transformations to the existing simulated map, rather than simulating the interloper emission from a simulated population of galaxies at the interloper redshift. This is an inaccurate representation of the large-scale structure of the interloper emission, but the stack is not very sensitive to large-scale structure, so this should not affect our results. The transformation was done using one of 11 rotations or reflections:
\begin{enumerate}
    \item One of three rotations (by $90^\circ$, $180^\circ$, or $270^\circ$) in the spatial axes.
    \item One of two reflections (in either the RA or Dec axis).
    \item A reflection across the frequency axis.
    \item A reflection across the frequency axis, combined with any of the spatial rotations or reflections.
\end{enumerate}
We then scaled the brightness of this transformed map to match the expected luminosity of foreground lines compared to CO. A factor of ten roughly corresponds to the difference between CO(1--0) at cosmic noon and the next most significant source of emission in our frequency range: CO(2-1) at $6 < z < 8$, which is expected to be roughly an order of magnitude fainter \citep{chung2024_globalsignals}. We then added the interloper map to the map of simulated CO emission.

Finally, we subtracted from the map a mean across the spatial dimensions in each spectral channel and across the spectral axis in each spaxel. This emulates the high-pass spatial and spectral filters in the actual COMAP data pipeline, which reject continuum emission almost entirely \citep{lunde2024_COMAPS2_PaperI}. For this reason, we ignored continuum emission when inserting an interloper signal.

\subsection{Simulated galaxy catalogue}\label{sec:simpipeline:simgalaxycatalogue}
We generated a galaxy catalogue from the set of simulated halos simply by logging the three-dimensional positions (and, optionally, luminosities) of some subset of the DM halos. This subset was generated by first cutting on luminosity, including only halos above a certain tracer luminosity $L_\mathrm{Ly\alpha,cut}$, in order to simulate observational completeness. We note that this should more properly be a cut in flux, as flux (rather than luminosity) is the quantity observed by the telescope. The conversion between flux and luminosity is redshift-dependent, so the completeness limit should also be redshift dependent. However, the effect of this change on the returned number counts of catalogued galaxies even at the edges of the COMAP redshift range is small (roughly 10\%). Additionally, most spectrographs have sensitivities that vary with wavelength, so the limiting catalogue luminosity will in practice be a function of redshift due to instrumental effects. Here, however, we treated it as constant.

We then selected $N_\mathrm{obj}$ halos with $L_\mathrm{Ly\alpha} > L_\mathrm{Ly\alpha,cut}$ randomly, weighting linearly by $L_\mathrm{Ly\alpha}$ as brighter objects are more likely to be detected. $N_\mathrm{obj}$ was set by the observational parameters of the galaxy survey being approximated. In practice, this selection is subject to the observing pattern of the galaxy survey used, but the stack is not sensitive to large scales, and thus large-scale spatial correlations should not be important here. Figure \ref{fig:joint_sim_slices} shows an example simulation of the Ly$\alpha$ population as well as the halos that are selected by the detectability cuts to be included in the mock catalogue.

To explore their effects, we also applied some other modifications to the galaxy catalogue:

\textit{Tracer velocity uncertainties}:~To simulate astrophysical velocity offsets that can occur between different spectral lines emitted by a galaxy, due to, for example, bulk inflows and outflows as well as redshift uncertainties due to varying spectral resolution of galaxy surveys, we optionally applied a scatter $\sigma_v$ to the catalogued redshift. We note that bulk velocity offsets may also be present between the two tracers, but assume that these will be well characterized and thus easily accounted for.

\textit{False positives:}~Especially at low signal-to-noise (S/N) ratios, galaxy surveys can occasionally misidentify either noise peaks or other emission lines originating from foreground or background galaxies as the spectral line of interest. When this happens, false positive detections occur, creating catalogue entries that do not correspond to actual galaxies in the target redshift range. We simulated this effect by generating catalogue entries with random spatial positions and redshifts and inserting these spurious entries into the catalogue. We quantified how many false positives are present in the simulated catalogue using a fraction, $f_\mathrm{FP}$, of total entries.

\begin{table}[]
\caption{\label{tab:LIM_params} Default values used in the generation of simulated LIM maps.}
    \centering
    \begin{tabular}{cc}
    \hline
   \multicolumn{2}{c}{\textit{LIM instrument parameters}} \\
   \hline
       $T_\mathrm{sys}$  & 44 K \\
        $N_\mathrm{f}$ & 19 \\
        $\Delta_\nu$ & 8 GHz (1028 Mpc)\\
        $\Delta_x$ & 4 deg (218 Mpc)\\
        $\delta_\nu$ & 31.25 MHz (3 Mpc)\\
        $\delta_x$ & $2$ arcmin (3.6 Mpc)\\
        \thetab & $4.5$ arcmin (8.2 Mpc)\\
        $\tau$ & 29 000 h \\
    \hline
    \multicolumn{2}{c}{\textit{LIM modelling parameters}}\\
    \hline
        CO $L(M_\mathrm{h})$ & Chung et al.~(2022) \\
        $A$ & $-2.85$ \\
        $B$ & $-0.42$ \\ 
        $\log C$ & $10.63$ \\
        $\log\frac{M}{M_\odot}$ & $12.3$ \\
        $\sigma_\mathrm{CO}$ & $0.42\ \mathrm{dex}$ \\
        Line broadening method & \vmp\ (Behroozi et al.~2019) \\
    \hline
    \multicolumn{2}{c}{\textit{Catalogue modelling parameters}} \\
    \hline
        Ly$\alpha$ luminosity function & Ouchi et al.~(2020) \\
        $L^*_\mathrm{Ly\alpha}$ & $8.49\times10^{42}\ \mathrm{erg\ s^{-1}}$ \\
        $\phi^*_\mathrm{Ly\alpha}$ & $3.90\times10^{-4}\ \mathrm{Mpc^{-3}}$ \\
        $\alpha$ & $-1.8$ \\
        $\sigma_\mathrm{Ly\alpha}$ & $0.41\ \mathrm{dex}$ \\
        $N_\mathrm{obj}$ & 1000 \\ 
        $L_\mathrm{Ly\alpha,cut}$ & $8.26 \times 10^9 L_\odot$\\
        $f_\mathrm{FP}$ & $0\%$ \\
   \hline
   \multicolumn{2}{c}{\textit{Joint parameters}} \\ 
   \hline
        $\rho$ & $-0.5$ \\
    \hline 
    \end{tabular}
    \tablefoot{Values are based on the COMAP Pathfinder experiment and the HETDEX catalogue of LAEs. Where applicable, distances are comoving distances calculated at the average LAE redshift of 2.8.}
\end{table}

\section{Three-dimensional stacking}\label{sec:stacking_methods}
\subsection{Summary of stacking methodology}\label{sec:stacking_methods:summary}

\begin{figure*}
    \centering
    \includegraphics[width=0.98\linewidth]{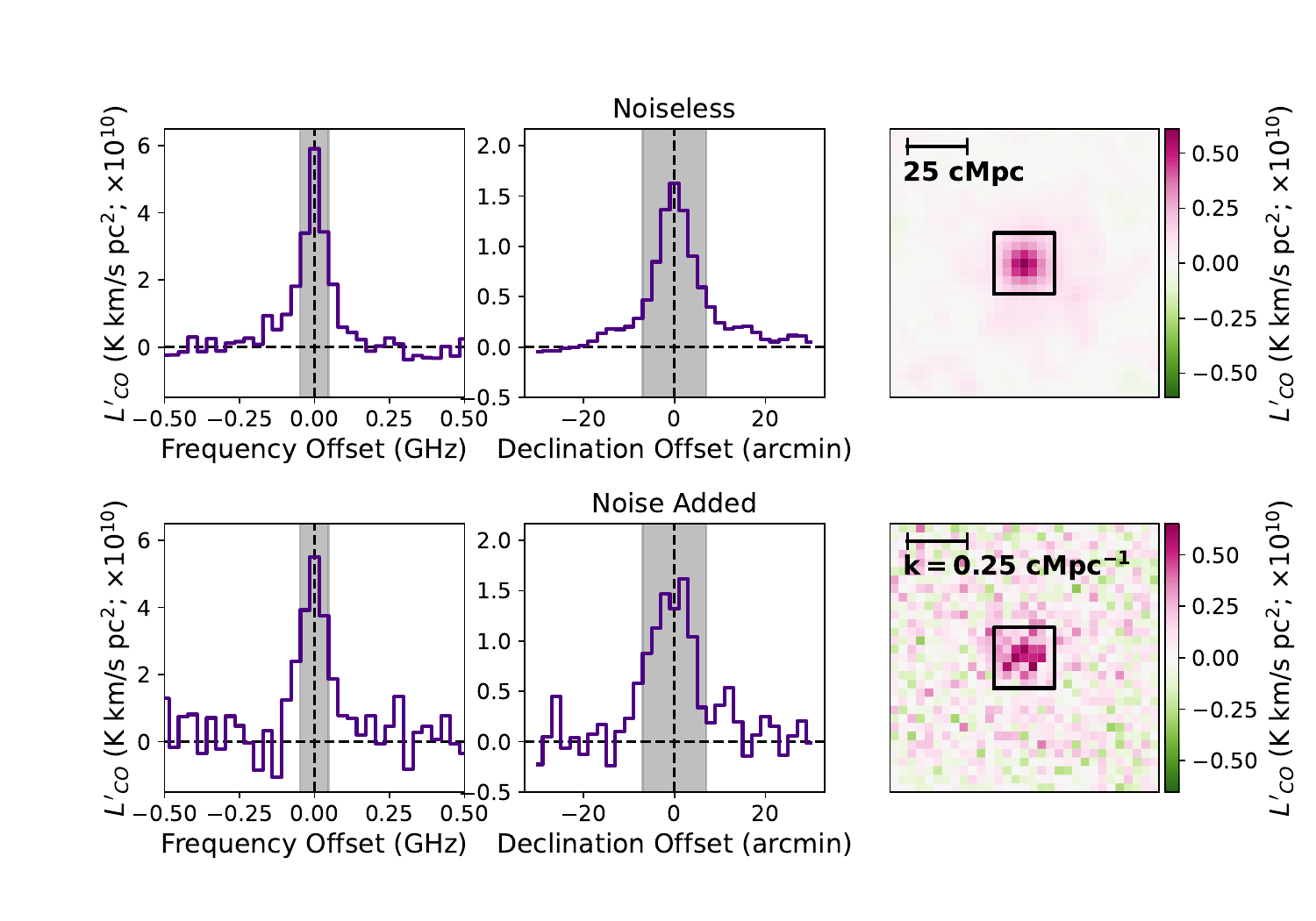}
    \caption{One- and two-dimensional projections of a simulated stack cubelet generated using the default simulation parameters. The top row shows a simulation realisation with no added noise (variations are due to individual simulated halos), and the bottom row shows the same simulation realisation with added radiometer noise (following \S\ref{sec:simpipeline:simmap}). \textit{Left:}Frequency spectrum of the central stack aperture. The $N_\mathrm{chan}$ frequency channels that are integrated over to generate the final stack luminosity are highlighted in grey. \textit{Center:} Spatial profile of the stack determined by summing over the three central frequency channels (those highlighted in the spectrum) into an image (shown in the right panel) and then collapsing the RA axis by summing over the three central spaxels. The spatial profile plots this quantity as a function of the angular offset in the declination direction from the stack centre. As in the spectrum, the width of the spatial aperture over which the emission is integrated to generate the final stack luminosity is highlighted in grey. \textit{Right:} Two-dimensional image with the $N_\mathrm{spax}\times N_\mathrm{spax}$ spatial aperture boxed in black.}
    \label{fig:default_stack}
\end{figure*}

We primarily followed the methodology for three-dimensional stacking established in \cite{dunne2024_ebossstacking}, and we refer the reader to this work for the full details of the stacking pipeline. In essence, the stack is a simple co-addition -- three-dimensional cutouts are taken from the LIM data cube at the angular and line-of-sight positions of objects included in a galaxy survey, and the cutouts are averaged together using inverse-variance weighting. The resulting stacked `cubelet' contains the average distribution of CO flux around the halos included in the Ly$\alpha$-based galaxy catalogue. Finally, the cubelet is summed over a square central aperture of $N_\mathrm{spax}^2$ spaxels (spatial pixels) in the spatial axes to create a one-dimensional spectrum. This spectrum is integrated over $N_\mathrm{chan}$ spectral channels to yield an average luminosity measurement of the regions that correspond to a catalogue object and are included in the stack. A stack performed on simulation realisations generated using the default parameters (listed in Table \ref{tab:LIM_params}) is shown in Figure \ref{fig:default_stack}.

Unlike the $N_\mathrm{spax}=3$ value used in \cite{dunne2024_ebossstacking}, in this work we took $N_\mathrm{spax} = 7$ ($\sim 25.9\ \mathrm{Mpc}$, or $k_\perp \sim 0.25\ \mathrm{Mpc}^{-1}$) as the default. In \cite{dunne2024_ebossstacking}, we defaulted to the $3\times 3$ aperture, as this is the nearest odd number to $1.5\times$ the FWHM of the beam using the COMAP beam width and spatial pixelisation. With a $3\times 3$ aperture, a catalogue object can be located anywhere within the central spaxel and still have its entire beam FWHM fall into the $N_\mathrm{spax}\times N_\mathrm{spax}$ aperture. However, as we discuss in detail below, we find that larger-scale clustering contributes extensively to the stacked signal, and a $7\times 7$ aperture actually maximizes the stack S/N. We also differed in our choice of $N_\mathrm{chan}$ -- here, we chose $N_\mathrm{chan} = 3$ ($\sim 12.3\ \mathrm{Mpc}$, or $k_\perp \sim 0.51\ \mathrm{Mpc}^{-1}$). We chose this after inspecting the default stacked spectrum (e.g.~Figure \ref{fig:default_stack}), which on average has a stack FHWM of $835\ \mathrm{km/s}$ (83.6 MHz). Three spectral channels ($936\ \mathrm{km/s}$, or 93.75 MHz) is the nearest odd number of channels to this width, and encompasses nearly all the flux from the stacked spectral line. This is narrower than the expected value in \cite{dunne2024_ebossstacking}, because that stack was performed on a catalogue of quasars with a known large velocity uncertainty. In the case of a galaxy catalogue with no significant uncertainty in redshift, the linewidth is much narrower (we investigate these effects further in \S\ref{sec:simresults:catalogue:zuncert}).

\subsection{Formalism for stack sensitivity}\label{sec:stacking_methods:stackmath}

As the goal of this work is to explore what experimental design and analysis parameters are optimal for the purposes of the stack, we attempt here to establish an analytical formalism for the stack sensitivity. We present this in terms of the stack detection S/N ratio, as we explore factors that impact both the signal and the uncertainty in the final stack. We write this ratio as 
\begin{equation}
    \frac{S}{N} = \frac{\langle L_\mathrm{LIM}\rangle}{\sigma_\mathrm{stack}},
\end{equation}
where $\langle L_\mathrm{LIM}\rangle$ is the average luminosity in the LIM tracer across the stacked regions (the target signal), and $\sigma_\mathrm{stack}$ is the uncertainty in that average. 

The nature of the stack means that the signal $\langle L_\mathrm{LIM} \rangle$ is a biased tracer of the overall CO signal at the target redshifts. Because the average in the stack is only over cutouts of the LIM cube centred around DM halos included in the galaxy catalogue, a region of space is only included in the average if it contains a halo emitting a detectable amount of (in this case) Ly$\alpha$ emission. We represent this effect with a `detectability' factor, which we represent as $S\left(L_\mathrm{Ly\alpha}\left(M_{\mathrm{h}},\rho\right)\right)$. This is a delta function (either a region of space is included or it is not) set by factors including the minimum luminosity detectable by the galaxy survey ($L_\mathrm{Ly\alpha, cut}$), the mass of the halo, the size of the region of space cut out for each catalogue object, and the correlation (or anti-correlation) between the scatter in the catalogue tracer luminosities and the LIM tracer luminosities of each halo, which we parametrize with $\rho$ (\S\ref{sec:simpipeline:correlated_scatter}).

Additionally, because LIM experiments are of relatively low spatial/spectral resolution (by design) compared to the galaxy survey from which the catalogue is derived, the region of space surrounding the catalogued object included in the stack is quite large. This means it is almost certain there will be neighbouring galaxies included in the $N_\mathrm{spax}\times N_\mathrm{spax}\times N_\mathrm{chan}$ central aperture, so $\langle L_\mathrm{LIM}\rangle$ is not a direct average over the catalogued galaxies themselves. Instead, there is another term contributing to $\langle L_\mathrm{LIM}\rangle$, which is the sum over the luminosities $L'_\mathrm{LIM}(M_\mathrm{h})$ of the $N(M_{\mathrm{h},i})$ neighbouring galaxies for each $i^\mathrm{th}$ catalogued galaxy. Because we are only probing the regions surrounding objects in a galaxy catalogue, the neighbouring galaxies are also not truly representative of the global population of DM halos (they are only included in the stack if they are adjacent to a catalogued halo, and thus are biased via proximity to the catalogue-selected objects). 

Overall, this results in a model for the stack luminosity that can be written as 
\begin{equation}\label{eqn:stack_avg_luminosity}
\begin{split}
    \langle L_\mathrm{LIM} \rangle = \frac{1}{N_h} \sum_{i} & S(L_\mathrm{Ly\alpha}(M_{h,i},\rho)) \times \\
    & \left[ L'_\mathrm{LIM}(M_{h,i}) + \sum_{n=0}^{N(M_{h,i})} L'_\mathrm{LIM}(M_{h,n}) \right].
\end{split}
\end{equation}
In other words, the stack luminosity is sensitive to both the LIM tracer luminosity of the objects included in the galaxy catalogue, and the degree to which those objects trace larger-scale overdensities of emitters in the LIM tracers. Both a catalogue of bright emitters or a highly biased catalogue will thus in theory return higher stack luminosities. Each of these parameters is tunable using the simulation setup described in \S\ref{sec:simpipeline}, and we explore the extent to which these different factors affect the overall significance of the stack in the following sections.

The uncertainty $\sigma_\mathrm{stack}$ is somewhat more straightforward. The stack is an inverse-variance weighted average, so assuming constant noise across the LIM map (which we do assume for the simulations presented in this work, although this is not true in general), the sensitivity in the stack relates to the sensitivity in an individual cutout's stack aperture as 
\begin{equation}
    \sigma_\mathrm{stack} = \frac{\sigma_\mathrm{aper}}{\sqrt{N_\mathrm{obj}}},
\end{equation}
where $N_\mathrm{obj}$ is the number of catalogue objects included in the stack. The signal in each stack aperture is from a sum across voxels, so using error propagation and again assuming a constant root mean square (RMS) noise value across the LIM cube, this can be written in terms of the noise level in each individual voxel (3D pixel), $\sigma_\mathrm{vox}$, as
\begin{eqnarray}
    \sigma_\mathrm{stack} &=& \frac{1}{\sqrt{N_\mathrm{obj}}} \sqrt{\sum_{N_\mathrm{chan}} \sum_{N_\mathrm{spax, x, y}}  \left(\sigma_\mathrm{vox}^2\right)} \\
    &=& \sqrt{\frac{N_\mathrm{chan} N_\mathrm{spax}^2}{N_\mathrm{obj}}} \sigma_\mathrm{vox}.    
\end{eqnarray}
In this case (using COMAP as an example), $\sigma_\mathrm{vox}$ is the radiometer noise described in Equation \ref{eqn:comap_radiometer}. The uncertainty is thus written as 
\begin{equation}\label{eqn:stack_sensitivity}
    \sigma_\mathrm{stack} = T_\mathrm{sys} \sqrt{\frac{N_\mathrm{chan} N_\mathrm{spax}^2 \Delta_x^2 }{N_\mathrm{obj}2\delta_\nu \tau  \delta_x^2 N_\mathrm{f} }}.
\end{equation}
This means that, in addition to the usual parameters that affect the sensitivity of a LIM experiment (such as the system temperature, the integration time, the number of feeds, etc.), which we do not explore here, the stack uncertainty is specifically dependent on the number of voxels included in the stack aperture in all three axes, as well as the voxel size in all three axes, and the number of objects included in the stack. 

Finally, we note that this uncertainty calculation only accounts for thermal noise in the COMAP data, and leaves out sample variance. \cite{chen2025_meerklassstack} have shown that the sample variance is an important consideration when calculating cosmological parameters from a stack. However, as this work is framed primarily in terms of detecting signal from a stack, we ignore the contribution to the uncertainty from the sample variance, as thermal noise should be the main driver of what is detectable with a given instrument and spectral line. The S/N ratio values listed throughout this work can be thought of as `detection significances'.

\section{Simulated stack results}\label{sec:simresults}

We explore the effects of varying different facets of both the LIM map and the galaxy catalogue on the stack sensitivity. We divide these effects into choices made while generating the stack itself (\S\ref{sec:simresults:stackparams}), factors in the experimental design of either the galaxy catalogue or the LIM experiment (\S\ref{sec:simresults:catalogue} and \S\ref{sec:simresults:map}), astrophysical factors affecting the LIM or the galaxy catalogue tracer spectral line (\S\ref{sec:simresults:astrophysics}), and interloper contamination in either experiment (\S\ref{sec:simresults:interlopers}). 

\subsection{Stack parameters}\label{sec:simresults:stackparams}

Some of the most obvious things that can be varied are in the choices made for the stack itself. Here, we test the size of the central aperture -- the number of pixels in the spatial ($N_\mathrm{spax}$) and spectral ($N_\mathrm{chan}$) directions that are summed over to determine the total luminosity of the stacked cubelet. Under the stacking methodology we use here, these must be an odd integer, and they should be as small as possible (as $\sigma_\mathrm{stack} \propto \sqrt{N_\mathrm{chan} N_\mathrm{spax}^2}$) while still capturing as much as possible of the flux from the stack.

\subsubsection{Spatial aperture size}\label{sec:simresults:stackparams:spatial}

\begin{figure}
    \centering
    \includegraphics[width=0.95\linewidth]{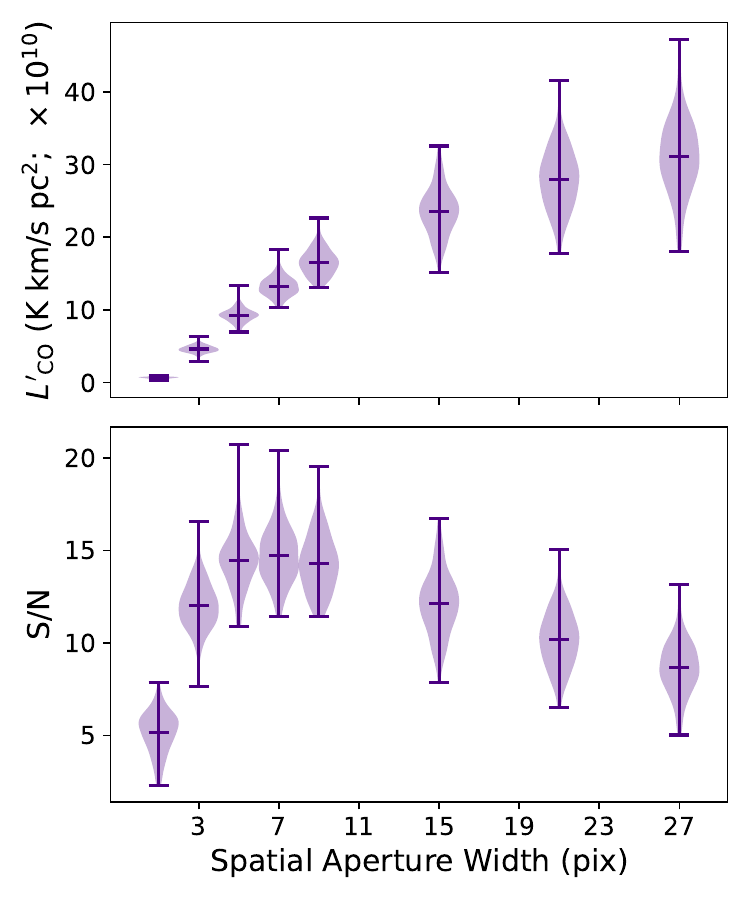}
    \caption{Violin plot showing how the luminosity (top) and the S/N (bottom) of the stack changes as the central aperture being summed over to determine the final stack luminosity is widened in the spatial axes. The default configuration uses a $3\times3$-spaxel aperture.}
    \label{fig:spatial_aperture}
\end{figure}

Here, we tested the various choices of spatial aperture size to see which returns the best stack S/N. We performed stacks on 99 different simulation realisations using the default configuration. For each, we determined the stack luminosity and S/N using square apertures with side lengths varying from one spaxel to nine spaxels, in steps of two (holding the aperture width constant in the spectral axis) as well as larger apertures of side lengths 15, 21, and 27. We plot the results in Figure \ref{fig:spatial_aperture}.

We find that the choice of central aperture size that maximizes the stack S/N is $N_\mathrm{spax} = 7$, although $N_\mathrm{spax} = 5$ and $N_\mathrm{spax} = 9$ provide very comparable results. This is contrary to the logic laid out in \cite{dunne2024_ebossstacking}, where we had chosen $N_\mathrm{spax} = 3$ so that a single object located anywhere inside the central spaxel will have the majority of its beam FWHM fall into the central aperture, while simultaneously integrating in as little noise as possible. While the noise in the stack does increase as $\sigma_\mathrm{stack} \propto N_\mathrm{spax}$, increasing the aperture size to $7\times7$ integrates in so much more luminosity that the S/N actually increases. The stack luminosity continues to increase considerably with increasing aperture size, although it flattens out in the larger apertures. This reflects the spatial profile of the stack luminosity, which has large wings extending out to $\sim 20$ arcmin (Figure \ref{fig:default_stack}). As the aperture gets larger than $\sim 7\times7$ the effects of the increased $N_\mathrm{spax}$ dominate the S/N, which begins to decrease again even as the luminosity continues to increase.

\begin{figure}
    \centering
    \includegraphics[width=0.95\linewidth]{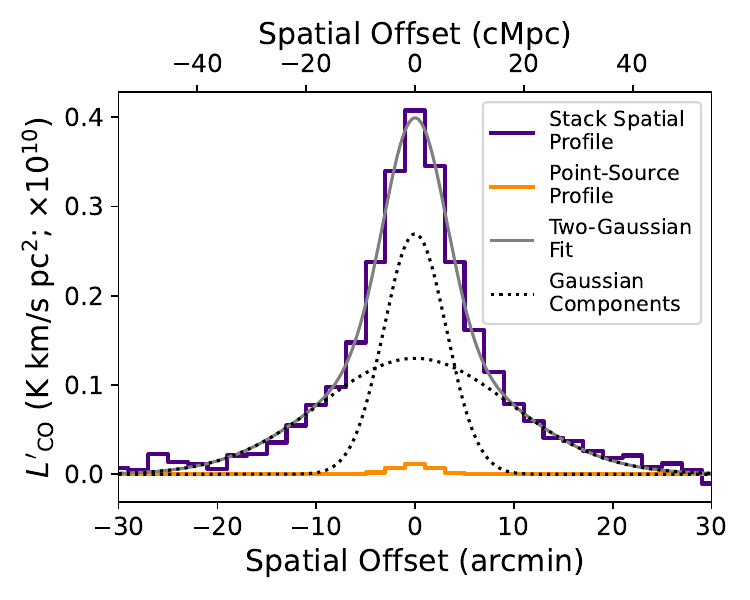}
    \caption{Average one-dimensional spatial profile of 99 noiseless simulated stack realisations. A double-Gaussian fit to the profile is shown in grey, with each of the Gaussian components shown as black dotted lines. The orange profile is a stack performed on simulations that have CO luminosity painted only onto the halos actually included in the mock galaxy catalogue, to demonstrate the extent to which the stacked signal is the result of halo clustering.}
    \label{fig:spatial_profile_fits}
\end{figure}

The fact that there is so much more luminosity at wider radii in the stack suggests that the stack is not well-explained by point sources in the central spaxel (i.e.~the objects actually included in the galaxy catalogue alone cannot account for all of the stack signal). We plot the average spatial profile of 99 noiseless simulations in Figure \ref{fig:spatial_profile_fits}. We also show a two-component Gaussian fit to this profile, including both a brighter, narrower component and a broader, dimmer component. This is motivated by the logic that the profile would be dominated by luminosity from the catalogued object and its immediate surroundings, and larger-scale cosmological clustering would also contribute. The profile is fit well with two Gaussian profiles (reduced $\chi^2 = 0.09$). The standard deviations of the two Gaussian components are $3.2 \pm 0.1'$ and $9.6 \pm 0.5'$, respectively.

To confirm that the stack signal is the product of larger-scale clustering, and not extended signal from the catalogued objects themselves, we also generate the profile of a stack on only the catalogued DM halos. We generate noiseless simulation realisations with the $L_\mathrm{CO}$ of all halos but those included in the mock galaxy catalogue artificially set to zero, and perform the stack on these simulation realisations. The resulting stack is also shown in Figure \ref{fig:spatial_profile_fits}. We find that not only is the spatial extent of this stack considerably reduced, but the luminosity of the stack itself is nearly negligible compared to the luminosity of the stack on the full mock LIM data. This suggests that the stack signal is so extended because, at least when using the default models for both CO and Ly$\alpha$, the signal is coming almost entirely from halos surrounding the catalogue object, rather than from the catalogue object itself. We explore the extent of this clustering in Section \ref{sec:discussion:clustering}, and its dependence on model choices in Section \ref{sec:discussion:comodels}. Although somewhat outside of the scope of this paper, we explore how model choices affect the spatial profile of the stack specifically in Appendix \ref{app:spatial_extent}.

\subsubsection{Spectral aperture size}\label{sec:simresults:stackparams:spectral}

\begin{figure}
    \centering
    \includegraphics[width=0.95\linewidth]{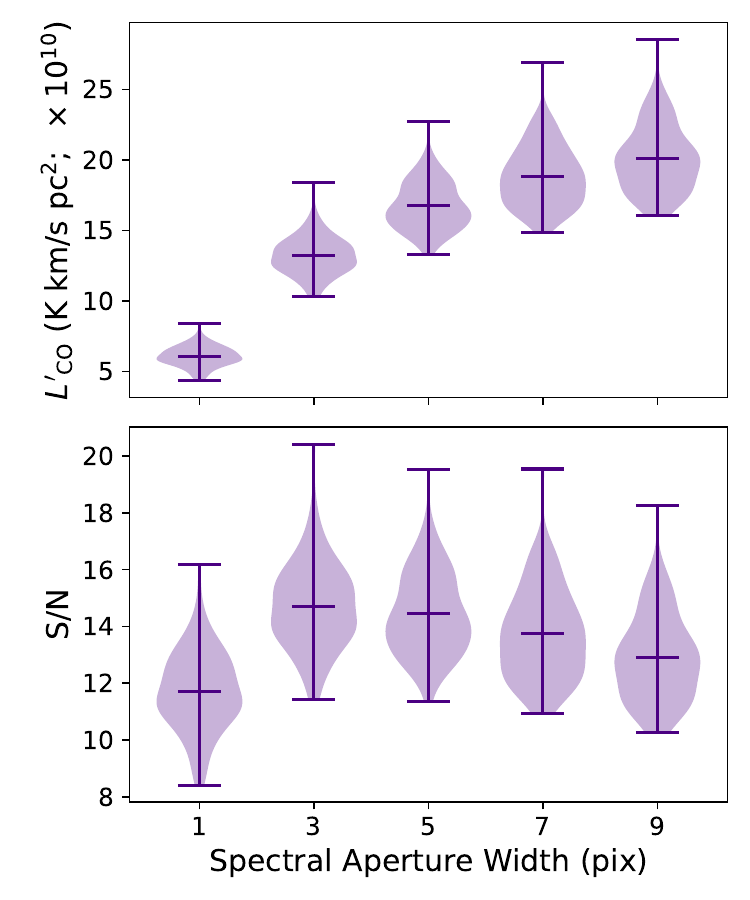}
    \caption{Violin plot showing how the luminosity (top) and the S/N (bottom) of the stack changes as the central aperture being summed over to determine the final stack luminosity is widened in the spectral axis. The default configuration uses a three-channel aperture.}
    \label{fig:spectral_aperture}
\end{figure}

We default to an aperture width of three channels in the spectral axis. Here, in addition to cosmological clustering, the stacked signal is broadened by astrophysical line broadening (rather than instrumental resolution). The default aperture width of three channels is chosen from visual inspection of the output stack spectrum, as astrophysical line broadening is not well characterized for CO at $z\sim 3$. As above, we test the choice of aperture width by performing stacks on 99 different simulation realisations, and extracting the central aperture luminosity using varying aperture widths. Here, we hold the spatial axes constant at $3\times3$ spaxels and vary only the spectral axis, stepping from one channel to nine channels in steps of two. The results are shown in Figure \ref{fig:spectral_aperture}.

In this case, we find that the three-channel aperture does indeed maximize the S/N of the stack. As with the spatial axes, the output stack luminosity continues to increase as the aperture gets wider. The noise increases as $\sigma_\mathrm{stack} \propto \sqrt{N_\mathrm{chan}}$, and begins to dominate after the three-channel aperture.

\begin{figure}
    \centering
    \includegraphics[width=0.95\linewidth]{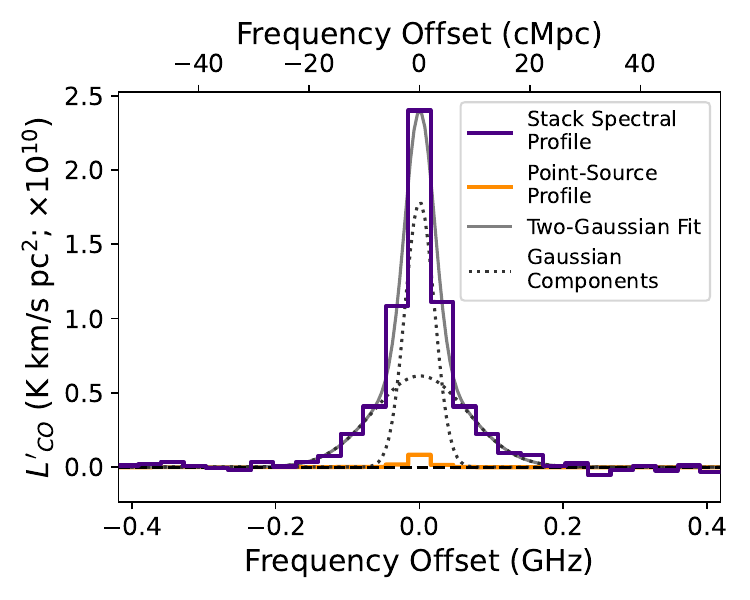}
    \caption{Average spectrum of 99 noiseless simulated stack realisations. A double-Gaussian fit to the profile is shown in grey, with each of the Gaussian components shown as black dotted lines. The orange profile is a stack performed on a simulated map with only the halos included in the Ly$\alpha$ catalogue assigned CO luminosities; all the other halos were given zero luminosity.}
    \label{fig:spectral_extent}
\end{figure}

As above, we tested the extent of the contribution from the objects actually included in the mock catalogue by comparing the spectrum of the stack to the spectrum of a stack performed on a map where only the sources included in the mock catalogue are assigned CO luminosity. This is shown in Figure \ref{fig:spectral_extent}. We fit a two-component Gaussian to the spectral profile of the stack. We find that there is also considerable clustering contributing to the width of the signal in the spectral axis. However, the spectral channels correspond to larger physical sizes than the spaxels do, so the three-channel aperture has roughly the same comoving size as the 7-spaxel aperture in the spatial directions.

\subsection{Galaxy catalogue experimental factors}\label{sec:simresults:catalogue}
As a resolved survey of galaxies is a much more established technique than LIM, there may potentially be several existing galaxy surveys to choose from when deciding to perform a stacking analysis with a LIM experiment. In this section, we explore how the makeup of the galaxy survey affects the stack result.

\subsubsection{Number of catalogue objects}\label{sec:simresults:catalogue:numobjects}

The main factor driving the sensitivity of the stack should be the number of catalogue objects being stacked, $N_\mathrm{obj}$, with the sensitivity improving as $\sqrt{N_\mathrm{obj}}$ for a case where the noise response is constant across the LIM map. To verify this hypothesis, we perform stacks of 100, 1000, and 10 000 catalogue objects on each of 99 different simulation realisations. For all other parameters, we use the `default' values listed in Table \ref{tab:LIM_params}.

\begin{figure}
    \centering
    \includegraphics[width=0.95\linewidth]{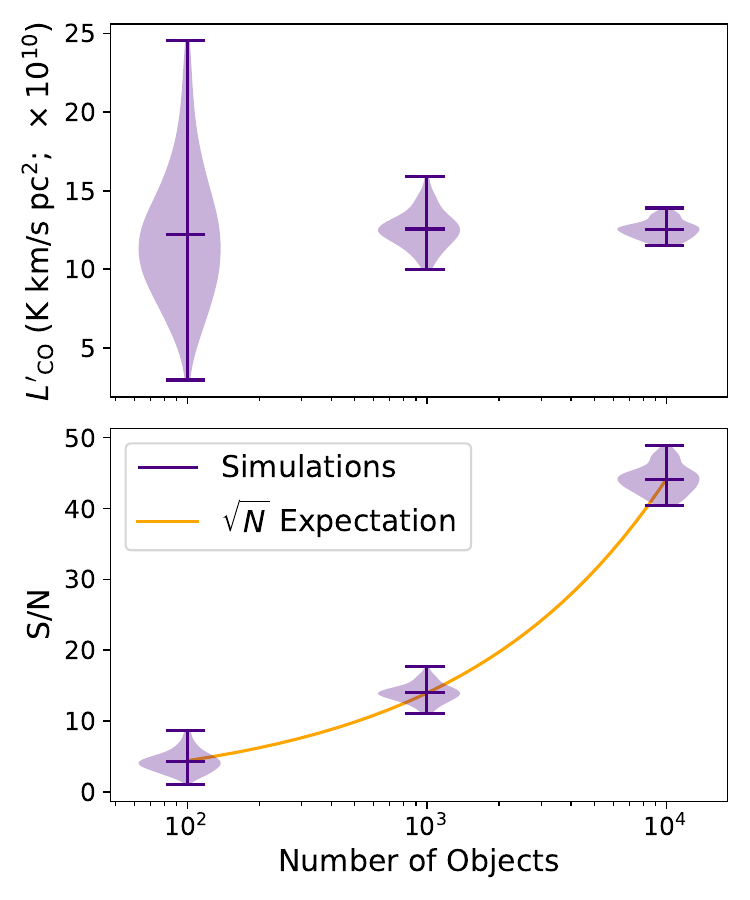}
    \caption{Violin plot showing how the luminosity (top) and the S/N (bottom) of the stack changes as more objects are included in the stacked catalogue. The orange line shows the $\sqrt{N_\mathrm{obj}}$ improvement, which is to be expected for a LIM map with only white noise.}
    \label{fig:number_objects}
\end{figure}

We plot the distribution of stack luminosity and S/N values for each $N_\mathrm{obj}$ in Figure \ref{fig:number_objects}. We also include for reference the theoretical $S/N \propto \sqrt{N_\mathrm{obj}}$ expectation for a stack with only white noise. Using the catalogue selection strategy we adopt (where catalogue objects can be selected above a cut-off luminosity, and the selection probability is linearly dependent on the luminosity of the catalogue object), the stacks with varying $N_\mathrm{obj}$ values all return extremely consistent luminosities. The variation of S/N with $N_\mathrm{obj}$ agrees well with theory.

\subsubsection{Redshift uncertainty}\label{sec:simresults:catalogue:zuncert}
Thus far (including in our derivation in \S\ref{sec:stacking_methods:stackmath}) we have assumed a galaxy catalogue with perfect redshift accuracy. However, this is unlikely to be the case in practice. Galaxy catalogues range from high-precision spectral surveys (such as DESI; \citealt{DESI_2024AJ....168...58D}) to photometric redshift surveys with large uncertainties (such as DES; \citealt{DES_2022PhRvD.105b3520A}). Additionally, while galaxy catalogues may have precise redshifts themselves, certain objects have inherent uncertain offsets between different redshift tracers \citep[for example, quasars can have large offsets between the optical lines used for eBOSS and molecular lines, due to inflows or outflows in the quasar host galaxy,][]{dunne2024_ebossstacking}. Often, spectral resolution is sacrificed in order to gain larger numbers of objects in a galaxy catalogue. 

\begin{figure}
    \centering
    \includegraphics[width=0.95\linewidth]{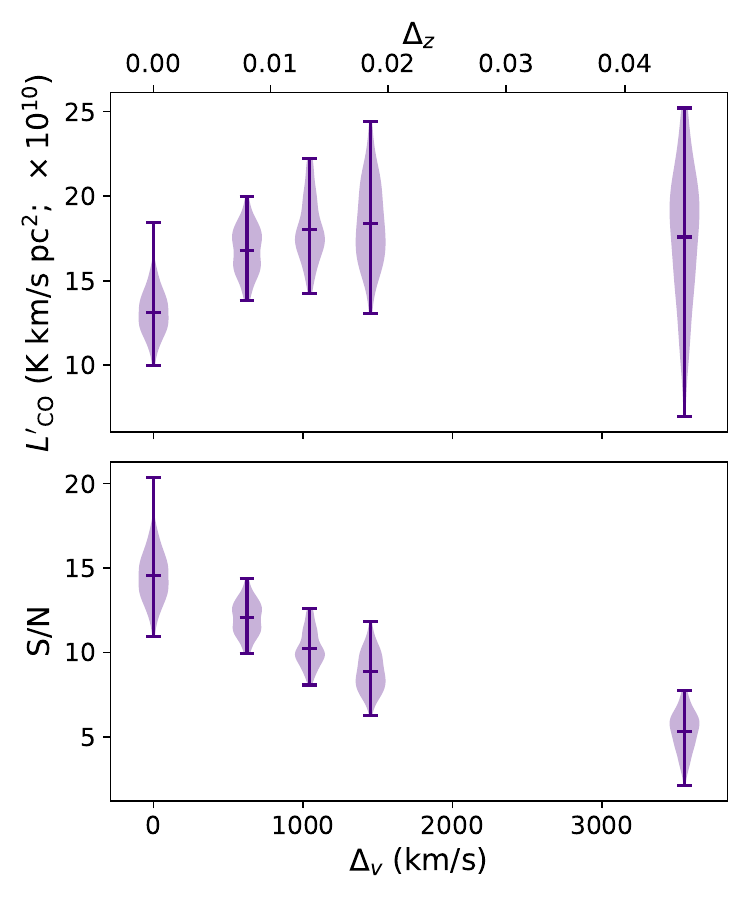}
    \caption{Violin plots showing the behaviour of the stack in returned luminosity (\textit{top}) and S/N (\textit{bottom}) with uncertainty, $\Delta_v$, in the redshifts of the galaxy catalogue being stacked on, assuming that uncertainty can be measured and the spectral aperture of the stack can be  widened to account for it. Each of these stacks are on 1000 catalogue objects. This analysis does not account for the increased numbers of objects available in some photometric catalogues.}
    \label{fig:redshift_uncert_violin_aperturecorrected}
\end{figure}

Because scatter in the redshift axis of a galaxy catalogue can typically be characterized, although not removed, it is possible to account for the redshift uncertainty when determining the spectral width of the stack aperture. We thus chose three uncertainty values corresponding to steps in the spectral width of the stack aperture. We measure the frequency width of a stack with no redshift uncertainty (this still contains some inherent width in its spectrum due to astrophysical line broadening), which we treat with a three-channel spectral aperture, and then scale this frequency width up to widths that match a 7-, 11-, and 15-channel aperture. We then convert this frequency width to a velocity width at the mean redshift of the catalogue objects ($z_\mathrm{mean} \sim 2.8$) and perform 99 different stacking realisations at each velocity uncertainty, matching the spectral aperture width to the line width for each stack. We plot the results in Figure \ref{fig:redshift_uncert_violin_aperturecorrected}. We repeat this analysis for a much larger redshift uncertainty approximating that of the Dark Energy Survey photometric catalogue at $z=1.2$, which is roughly $\Delta_z = 0.045$ \citep{DES_photz_2021ApJS..254...24S,DES_2022PhRvD.105b3520A}.

We find that collecting the signal scattered out of the original stack aperture by redshift uncertainty works very well -- the returned luminosity value for each of these stacks with added uncertainty is the same and is actually larger than the returned luminosity for the stack on the unscattered catalogue. This is because, as shown above, the linewidth of the stack is not actually a perfect Gaussian -- galaxy clustering and astrophysical line broadening, in addition to the redshift uncertainty, all contribute to widening the line. While the luminosity stays constant, the S/N ratio decreases considerably with increasing uncertainty, even for the smaller, spectroscopic levels of uncertainty. This is because this strategy adds additional spectral channels to the stack (i.e.~increases $N_\mathrm{chan}$), which adds to the noise as $\sqrt{N_\mathrm{chan}}$ (Equation \ref{eqn:stack_sensitivity}). In simpler terms, the signal in the stack is spread over a larger region of the noise floor. At the photometric redshift uncertainty $\Delta_z = 0.045$, the decrease in sensitivity is extremely significant -- the average S/N in the $\Delta_z = 0.045$ case is 5.3, only a third of the sensitivity of the unscattered case.

However, photometric galaxy catalogues also benefit from considerably increased source densities. We treat the DES Y6 Gold catalogue and the DESI ELG sample as examples of photometric and spectroscopic galaxy catalogues, respectively. This is merely an illustrative example, as neither catalogue has coverage in the redshift range we are simulating in the rest of this work. The DES Y6 Gold catalogue has approximately 28.9 confident galaxy detections per arcmin$^2$, or 104,040 deg$^{-2}$. In comparison, the DESI ELG sample has an on-sky density of roughly 2400 deg$^{-2}$ \citep{desi2024_sampledefinition} -- the DES photometric catalogue has a factor of 43.5 greater object density. This is more than enough to offset the attenuation due to redshift uncertainty, at least in this case. We find that a stack on a catalogue with a DES-like redshift uncertainty ($\Delta_z = 0.045$), but also a DES-like increase in object density (43,500 stack objects) returns a stack S/N ratio of 37.5. Photometric catalogues may then work very well for a stacking analysis, provided they are rich in available objects. To yield a stack detection at the same significance, a catalogue with the DES redshift uncertainty would need to have $7.5\times$ as many objects as a catalogue with no redshift uncertainty.

\subsection{LIM data experimental factors}\label{sec:simresults:map}
The resolution of various LIM experiments varies considerably in both the spatial and spectral axes, which affects the stack outcome. In this section, we test the effects of varying the spectral resolution, the pixelisation, and the beam width of the LIM maps themselves. 

\subsubsection{LIM spectral resolution}\label{sec:simresults:map:spectralresolution}
The spectral resolution of the map is particularly interesting, as there is a large variation in spectral resolution across existing and proposed LIM experiments, based on the spectral lines being pursued and the technologies being employed at the receiver front end. For example, CONCERTO \citep{concerto2020_intro}, a Kinetic Inductance Detector (KID) experiment with an absolute spectral resolution of $\sim 1.5\ \mathrm{GHZ}$ \citep{fasano2022_concerto}, FYST \citep[][KID-based with $R\sim 100$]{ccatprime2023_overview}, or TIME \citep[][using transition edge sensor bolometers with $R\sim 100$]{crites2014_timeSPIE, sun2021_time} all have velocity resolutions $(\Delta_\mathrm{v} \sim 1000-3500\ \mathrm{km\ s^{-1}})$. This is up to an order of magnitude broader than the default value used here ($\Delta_\mathrm{v} \sim 300$ \kms), for COMAP.
SPHEREx \citep[][]{dore2014_spherexintro,dore2016_spherexscienceI,dore2018_spherexscienceII}, using optical linear variable filters at $R=35-130$, will have a resolution ranging from $2300$ to $8500$~\kms. 

We vary the spectral resolution into which the maps are being binned while preserving the size of the central frequency aperture at 91.75 MHz. Because the current stacking methodology requires there to be an odd number of frequency channels in the central aperture, we place one, three, and five channels across the aperture, corresponding to channel widths of 91.75 MHz, 31.25 MHz (the COMAP science resolution), and 18.75 MHz. 
We create 99 different simulation realisations at each spectral resolution, and perform stacks on each realisation to measure the effects of changing spectral resolution on the sensitivity of the stack. In each case, we extract the central 91.75 MHz to calculate line luminosity values. Theoretically, broadening the spectral resolution while maintaining the same stack aperture width in frequency should not affect the stack sensitivity (from Equation \ref{eqn:stack_sensitivity}, $\sigma_\mathrm{stack} \propto \sqrt{N_\mathrm{chan} / \delta_v}$, so the two parameters balance each other out).

\begin{figure}
    \centering
    \includegraphics[width=0.95\linewidth]{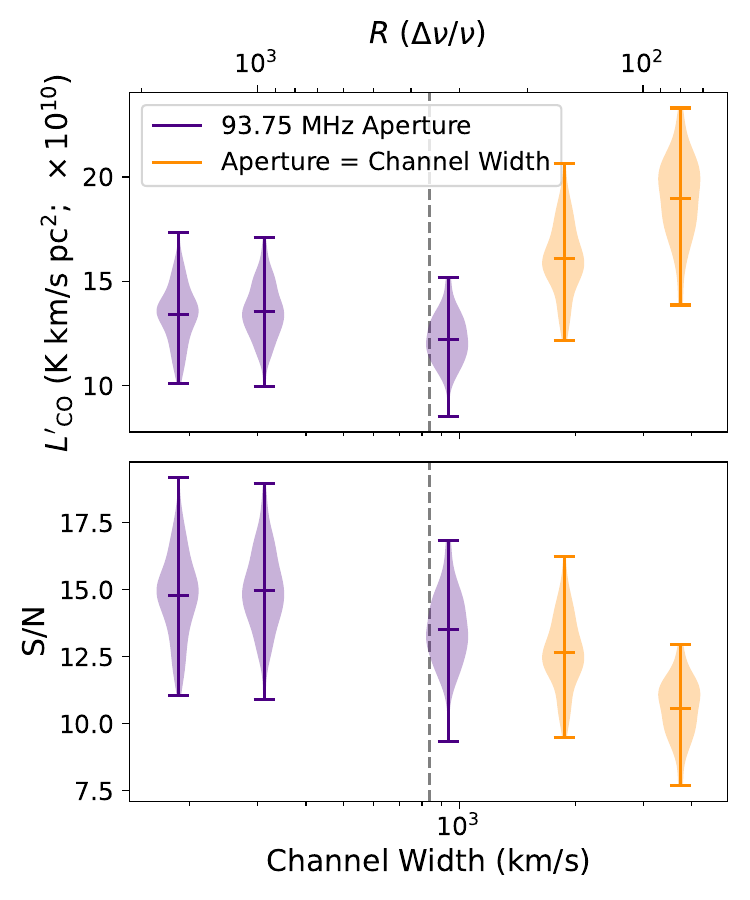}
    \caption{Violin plot showing the distributions of luminosity values (\textit{top}) and S/N ratios (\textit{bottom}) output from the stack for different simulation realisations at various LIM map spectral resolutions. Channel widths with indigo violins are those where the stack aperture width stays constant at 93.75 MHz as the channelisation changes. Orange violins have individual channels larger than 93.75 MHz -- their stack apertures are the size of a single spectral channel (the smallest possible size). The grey dashed line indicates the FWHM of the stack linewidth.}
    \label{fig:spectral_res_violin}
\end{figure}

Additionally, we perform stacks on each of 99 different simulation realisations at resolutions of 187.5 and 375 MHz (corresponding to roughly 1900 km/s and 3700 km/s, respectively). We do this to explore the effects of a LIM map with spectral resolution broad enough that a single frequency channel is much larger than the FWHM of the spectral line. These resolutions are wider than the chosen aperture size, so in each case we take the aperture size equal to the spectral resolution, and include only the central frequency channel in the stack aperture. Because the aperture size is now increasing, the assumption that the stack sensitivity should remain the same no longer holds. The results of these stacks are shown in Figure \ref{fig:spectral_res_violin}.

We find that when all other parameters (including the size of the stack aperture) are kept constant, improving the spectral resolution (decreasing the channel width) actually marginally improves the S/N ratio of the output stack, albeit with a large scatter between different simulation realisations. This trend is likely an artefact of the current aperture extraction strategy: currently, the cutouts of the LIM data corresponding to each catalogue object are only centred by lining up the central spectral channel. Thus, catalogue objects located anywhere within the central channel are treated identically. This introduces a velocity uncertainty the width of the central channel, in practice equivalent to convolving the average spectral profile of the objects being stacked with a boxcar function the size of the central channel. Shrinking the size of the channel shrinks the boxcar being convolved in, and retains more flux in the central stack aperture.

We note that this effect is mitigated by improving the spectral resolution of the LIM data, but it could equally be mitigated by exploring more optimal aperture extraction techniques, such as matched filtering \citep[e.g.][]{zubeldia2021_matchedfilters} or an interpolation-based centring. Still, even when more optimal techniques are used, a better LIM spectral resolution is likely to be advantageous for the stack, as it will put more data points across the stacked spectrum. This will enable a more accurate measurement of the average line profile of the objects included in the stack or make matched filtering or interpolation extraction techniques more effective.

In the case where the channel width is considerably larger than the stack FWHM (so the aperture width of the stack in the spectral axis is forced to increase), the overall measured stack luminosity increases, as more of the wings of the line emission are included in the central aperture. However, as the aperture size increases, the balance between $N_\mathrm{chan}$ and $\delta_v$ is no longer maintained in $\sigma_\mathrm{stack}$, so as expected the S/N ratio degrades. As the stack aperture gets wider, more regions of the spectrum that contain only noise are being integrated into the final stack, and the line luminosity increasing is not sufficient to counteract the effects of this increased noise. Experiments with larger spectral channels will have to account for this effect if they choose to perform stacking experiments.

\subsubsection{LIM spatial pixelisation}\label{sec:simresults:map:spatial_pix}

While the angular resolution (beam) of a single-dish LIM experiment is set by the size of the antenna being used and the illumination of that antenna by the feed(s), the pixelisation of the angular axes is decided in the map-making steps \citep[e.g.~][]{lunde2024_COMAPS2_PaperI}. As in the spectral axis, where the ability of the aperture extraction to concentrate signal under the current stacking methodology is limited by the width of the central frequency channel, it may be possible to improve the sensitivity of a stacking analysis by oversampling the beam and concentrating signal more effectively. 

We explore this possibility by varying the spatial resolution of the simulated maps, following a similar methodology to \S\ref{sec:simresults:map:spectralresolution}. Here, we hold the spectral resolution constant and instead vary the pixel size in the angular axes. Similar to the spectral analysis, we maintain the central aperture at a constant area and vary the number of spaxels placed across it. Again, the stack methodology requires an odd number of spaxels included in the central aperture in each angular dimension, so we place $1^2$, $3^2$, $5^2$, and $7^2$ spaxels across the area of the central aperture. These correspond to pixel scales of $6'$, $2'$, $1.2'$, and $0.857'$ per side. As above, $\sigma_\mathrm{stack} \propto \sqrt{N_\mathrm{spax}^2/\delta_x^2}$, so increasing the spectral resolution while maintaining the same aperture area should not affect the sensitivity of the stack.

\begin{figure}
    \centering
    \includegraphics[width=0.95\linewidth]{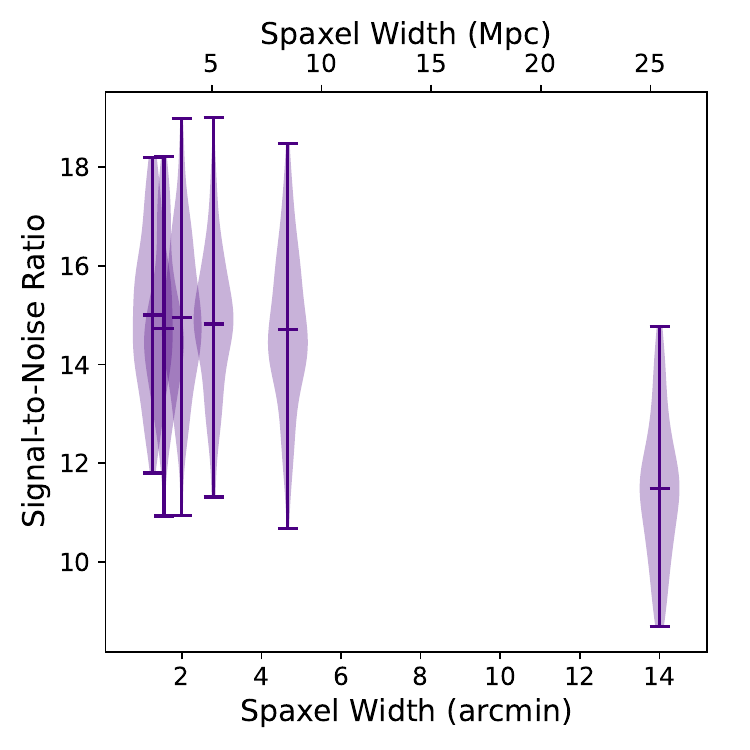}
    \caption{Violin plot showing the distributions of S/N ratios output from the stack as a function of the pixel size of the LIM data while holding the beam size constant.}
    \label{fig:spatial_res_violin}
\end{figure}

We create 99 different simulation realisations at each pixel size, performing stacks on each realisation. In each case, we extract a central $6'\times 6'$ aperture in the spatial axes. We show the S/N ratio of each stack as a function of its resolution in Figure \ref{fig:spatial_res_violin}. As with the spectral axis, improving the spatial pixelisation does marginally improve the stack S/N ratio, likely for the same reasons -- the catalogue object could be anywhere inside the central spaxel, so the average stack spatial profile is convolved with a 2D boxcar the width of the central spaxel. This effect is especially apparent in the case with $14'$ spaxels, where only the central spaxel is included in the stack -- objects located anywhere near the outskirts of the central spaxel will have almost half of their signal left out of the stack. Again, improved aperture extraction techniques could improve the concentration of signal as effectively as decreasing the pixel size.

\subsubsection{Beam size}\label{sec:simresults:map:spatial_beam}

Stacking analyses are likely to be far from the primary consideration setting the actual spatial resolution (in the case of COMAP, the beam size) of a LIM experiment. Nevertheless, for completeness, we explore the effects of the beam size on the stack sensitivity. We hold both the spatial and spectral pixel sizes constant, and vary the size of the beam being convolved into the synthetic LIM map.

\begin{figure}
    \centering
    \includegraphics[width=0.95\linewidth]{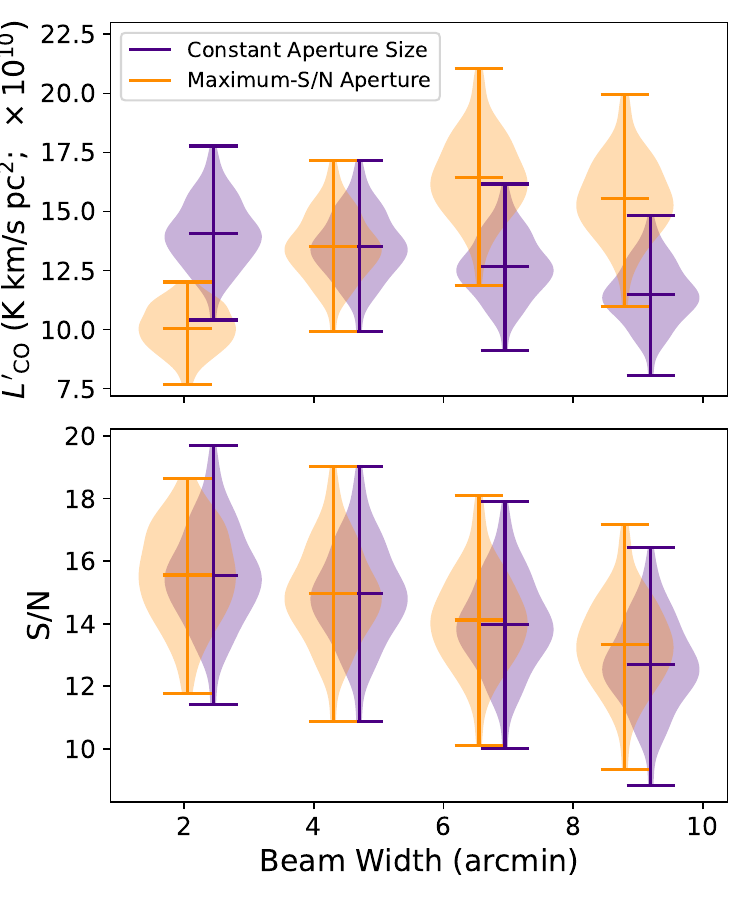}
    \caption{Violin plot showing the distribution of S/N ratios output from simulated stacks as a function of the beam FWHM (i.e.~angular resolution) of the LIM experiment while holding the pixelisation constant. We show a version where we hold the size of the spatial aperture constant as well as a version where we set the spatial aperture to the size that maximizes the S/N ratio of the stack. Violins are offset slightly in the x-axis for legibility.}
    \label{fig:beam_size_violin}
\end{figure}

We explore four steps in beam area -- we choose beam FWHM values of $2.25'$, $6.75'$, and $9.0'$ as well as the default value of $4.5'$. Because of the extended spatial distribution of the signal, it is difficult to scale the aperture size to the beam such that the same amount of signal falls into the central aperture at each beam size -- we have already shown that the stack cannot be treated as a point source. Instead, we compare the output S/N ratios of the stacks on different beam sizes in two separate ways. First we hold the aperture size constant at $N_\mathrm{spax} = 7$ in the spatial axes, and measure the output S/N ratio at each beam size. Second we determine the aperture size that maximizes the S/N ratio for each beam FWHM, and compare the maximum-S/N stacks against each other. For the $2.25'$ case the maximum S/N ratio is at $N_\mathrm{spax} = 5$, for the $4.5'$ case the maximum S/N ratio is at $N_\mathrm{spax} = 7$, and both the $6.75'$ and $9.0'$ cases have their S/N ratio maximized at $N_\mathrm{spax} = 9$. The results of these tests performed on stacks using 99 different simulation realisations at each beam FWHM are compared in Figure \ref{fig:beam_size_violin}.

We find that, in the case where the aperture size is held constant as the beam FWHM changes, the stack luminosity (and thus S/N ratio) does indeed decrease with increasing beam FWHM. Increasing the beam FWHM smears the signal more, pushing more of the luminosity from the stack outside the central aperture, so this is perhaps trivial. More interestingly, the stack S/N ratio falls off with beam FWHM at a nearly identical rate when the aperture size is chosen to maximize the S/N ratio. As the beam FWHM gets larger, the signal is being spread over a larger region of the noise floor. The aperture that maximizes the S/N ratio is balancing integrating in more luminosity from the outskirts of the stacked signal with the additional contribution of more noise as more spaxels are included in the central aperture. In any case, a more concentrated signal leads to higher S/N ratio.

We note that, while this analysis may seem very similar to \S\ref{sec:simresults:map:spatial_pix}, there is a key difference in the noise properties of the maps. In \S\ref{sec:simresults:map:spatial_pix}, where the pixelisation is being changed, the fraction of the observing time being spent in each spaxel and thus the per-spaxel noise properties vary across the different resolutions (i.e.~$\delta_x$ and $N_\mathrm{spax}$ are both changing). Thus, while smaller spaxel sizes sampled the beam better, they simultaneously had more noise per spaxel, so the two effects were working in concert. Here, $\delta_x$ is staying the same. The only variation is in the amount of signal spread outside the central aperture by the beam and potentially $N_\mathrm{spax}$.

\subsection{Astrophysical factors affecting the stack}\label{sec:simresults:astrophysics}

In addition to the experimental factors affecting the LIM data or the galaxy catalogue (which mostly come into the stack S/N ratio through the uncertainty in the stack, Equation \ref{eqn:stack_sensitivity}), many basic astrophysical factors will also affect the outcome of the stack. These will directly affect the measured stack luminosity and affect the S/N ratio as shown by Equation \ref{eqn:stack_avg_luminosity}. These factors tend to be considerably more nebulous than the experimental factors, as they have not yet been reliably measured at $z=3$, and, in the case of the CO and Ly$\alpha$ lines targeted here, are extremely difficult to model even in hydrodynamic simulations. Indeed, measuring these quantities is part of the goal of LIM experiments.

\subsubsection{Choice of LIM model}\label{sec:simresults:astrophysics:comodel}

A large source of uncertainty in the modelled stacks is the CO luminosity of DM halos at $z\sim 3$, $L_\mathrm{CO} (M_\mathrm{h})$. Although considerable effort has been made to accurately model these emitters on cosmological scales \citep[e.g.][]{lidz2011_co21cm,li2016_comodelling,padmanabhan2018_comodel,keating2020_mmime,chung2021_comapforecasts}, this is a phase space that has largely been unconstrained by observations. The best current constraints at these redshifts come from COMAP \citep{chung2024_COMAPS2_PaperIII} but the models that have not yet been excluded still predict luminosities that differ by an order of magnitude. We test three different models of CO emission, each described in \S\ref{sec:simpipeline:co_luminosity_sim}, to see how this affects the stacked signal. These models are C22 \citep[][the default model]{chung2021_comapforecasts}, P18 \citep{padmanabhan2018_comodel}, and L16 \citep{li2016_comodelling}. We chose not to explore any bulk changes to the CO luminosity, as these would increase the luminosity of all DM halos by the same factor and thus simply raise the stack luminosity by that same factor.

\begin{figure}
    \centering
    \includegraphics[width=0.95\linewidth]{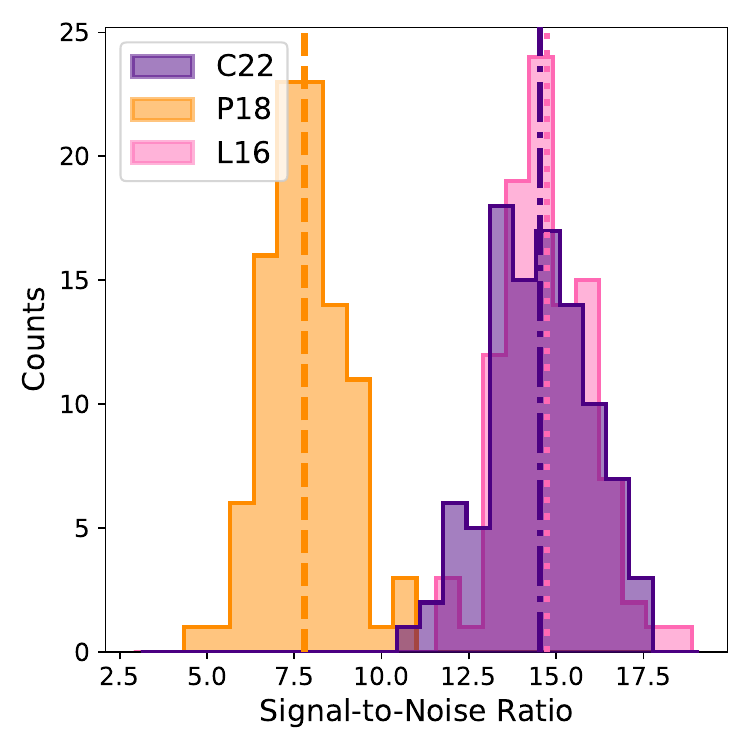}
    \caption{Histograms of the output S/N ratio for each model of the CO luminosity of a DM halo as a function of its mass, $L(M_\mathrm{h})$. Mean values for each model are shown as vertical lines. The three models are described in \S\ref{sec:simpipeline:co_luminosity_sim}.}
    \label{fig:co_funcs_comparsion}
\end{figure}

We ran 99 different simulation realisations for each model of CO emission, and compare the output S/N ratios for each in Figure \ref{fig:co_funcs_comparsion}. The C22 and L16 $L(M_\mathrm{h})$ functions yield very similar S/N ratios, while the P18 $L(M_\mathrm{h})$ function with $f_\mathrm{duty}=0.1$ has considerably lower significance. This is an interesting result, especially when compared against the $L(M_\mathrm{h})$ functions plotted in Figure \ref{fig:co_lum_funcs}. At most values of $M_\mathrm{h}$, the returned luminosity under the P18 function falls between the $L_\mathrm{CO}$ values from C22 and L16. L16 provides higher $L_\mathrm{CO}$ values at low $M_\mathrm{h}$, and C22 provides higher $L_\mathrm{CO}$ values at high $M_\mathrm{h}$. The cross-over is at roughly $1.5\times 10^{12}\ \mathrm{M_\odot}$, at which point all three $L(M_\mathrm{h})$ functions yield the same $L_\mathrm{CO}$. Based on this, one would naively expect either the C22 or L16 model to dominate, based on the halo mass selected by the catalogue luminosity function, and P18 to fall somewhere in the middle. We explore the implications of this more thoroughly in \S\ref{sec:discussion:comodels}.

\subsubsection{Choice of catalogue luminosity function}\label{sec:simresults:astrophysics:catalogue}

The returned luminosity of the stack does not depend directly on the Ly$\alpha$ luminosity of the halos. However, the $L_\mathrm{Ly\alpha}$ of the halos does determine which DM halos are actually included in the stack (via the selection function $S(L_\mathrm{Ly\alpha}(M_{h},\rho))$ in Equation \ref{eqn:stack_avg_luminosity}), which has the potential to significantly impact the stack outcome.

\begin{figure}
    \centering
    \includegraphics[width=0.95\linewidth]{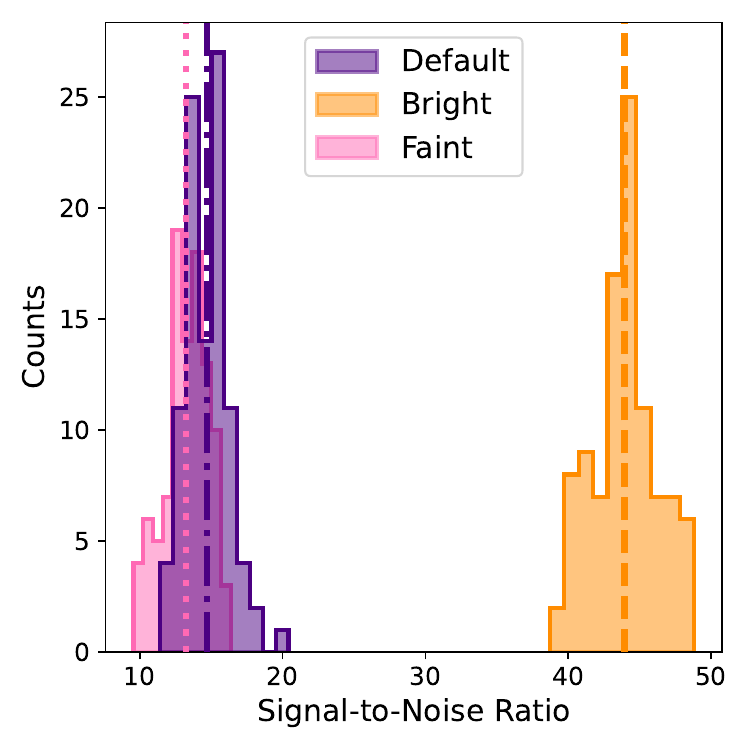}
    \caption{Histograms of the output S/N ratio for each model of the luminosity function of the emission line used in the galaxy catalogue being stacked on (in this case, Ly$\alpha$). The different catalogue models are described in \S\ref{sec:simpipeline:lya_luminosity_sim}.}
    \label{fig:lya_func_snrs_histograms}
\end{figure}

To test how the galaxy catalogue affects the behaviour of the stack we run 99 different simulation realisations for each of the Schechter function parameters described in Section \ref{sec:simpipeline:lya_luminosity_sim}. These include parameters fit to observations of LAEs at $z=3$ (the `default' model), parameters fit to quasars (which trace a more stochastic population of higher-mass halos; the `bright' model), and parameters fit to LAEs at $z=0.3$ (a luminosity function more dominated by the many low-mass halos with faint luminosities; the `faint' model). The output S/N ratios of stacks performed using each model are shown in Figure \ref{fig:lya_func_snrs_histograms}. Because we hold all other parameters constant, stacks on each model have the same noise levels -- differences in S/N ratio are driven solely by the output stack luminosity.

We find that the `faint' galaxy catalogue has properties that are very similar to the default LAE catalogue, while the `bright' catalogue is considerably brighter (by a factor of more than three). For all catalogue models, larger halo masses correspond to brighter luminosities in the catalogue tracer, so the bright catalogue model should be selecting only the largest DM halos. These will both be the brightest halos themselves and be the most biased (they will have more, and brighter, neighbouring halos contributing to the second term in Equation \ref{eqn:stack_avg_luminosity}). The combination of these effects should explain the extremely bright stack. We test this further in \S\ref{sec:discussion:comodels}.

\subsubsection{Correlation in scatter between tracer luminosities}\label{sec:simresults:astrophysics:rho}

Finally, because the stack is an average of the LIM tracer luminosities of the DM halos included in the galaxy catalogue, it will be affected by the halo-to-halo correlation between the LIM tracer luminosity and the catalogue tracer luminosity. As in \S\ref{sec:simresults:astrophysics:catalogue}, this is only a selection effect (coming in through the $S(L_\mathrm{Ly\alpha}(M_{h},\rho))$ factor in Equation \ref{eqn:stack_avg_luminosity}), and not a direct scaling of the luminosity values being averaged.

\begin{figure}
    \centering
    \includegraphics[width=0.95\linewidth]{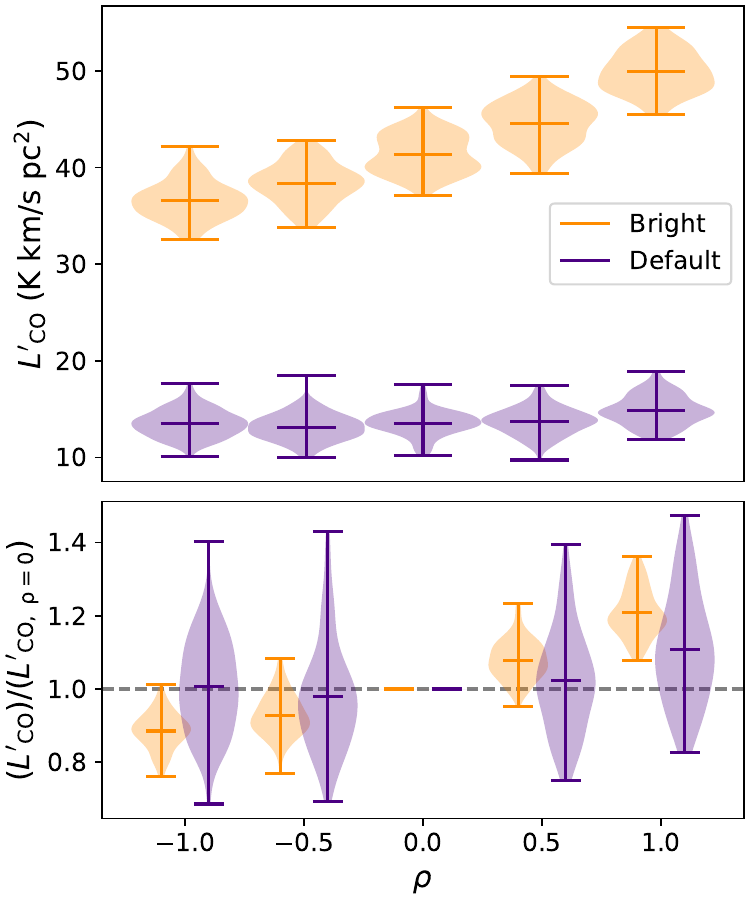}
    \caption{Effect of varying the correlation in the two luminosity values' scatter on the output stack luminosity. \textit{Top:} Violin plots showing the stack luminosity for each $\rho$ value for both the default choice of the Ly$\alpha$ luminosity function $z\sim 3$ and the bright AGN-like version. \textit{Bottom:} Output luminosity values normalised to the value at $\rho=0$ to show relative changes between values of $\rho$. Violins are offset slightly in the x-axis for clarity. }
    \label{fig:rho_luminosity_changes}
\end{figure}

We stack 99 simulation realisations at five different values of $\rho$ spanning from completely anti-correlated to completely correlated scatter in halo luminosities ($\rho = [-1, -0.5, 0, 0.5, 1]$). The output luminosity distributions of the resulting stacks are shown in Figure \ref{fig:rho_luminosity_changes}, both as absolute values and relative to the output luminosity for completely uncorrelated scatter ($\rho=0$). We repeat this exercise for the bright, AGN-like catalogue luminosity function, to explore interactions between the catalogue luminosity function and $\rho$ in determining how halos are selected and the resulting stack luminosity. 

We find that $\rho$ does have an effect on the returned luminosity, but this effect is very subdominant compared to the luminosity function of the catalogue tracer. Additionally, we find that varying $\rho$ has proportionally more of an effect on the output stack luminosity for the more stochastic, top-heavy `bright' catalogue luminosity function than for the flatter default one. We explore this behaviour in more detail in \S\ref{sec:discussion:comodels}.

\subsubsection{Astrophysical line broadening}\label{sec:simresults:astrophysics:linebroadening}

Astrophysical line broadening of the LIM tracer emission is likely to have an effect on the stack degenerate with redshift uncertainty in the galaxy catalogue -- it widens the signal in the spectral axis, thus possibly moving more signal outside the stack aperture. Unlike redshift uncertainty, this is inherent to the LIM tracer and not instrumental in origin.

\begin{figure}
    \centering
    \includegraphics[width=0.95\linewidth]{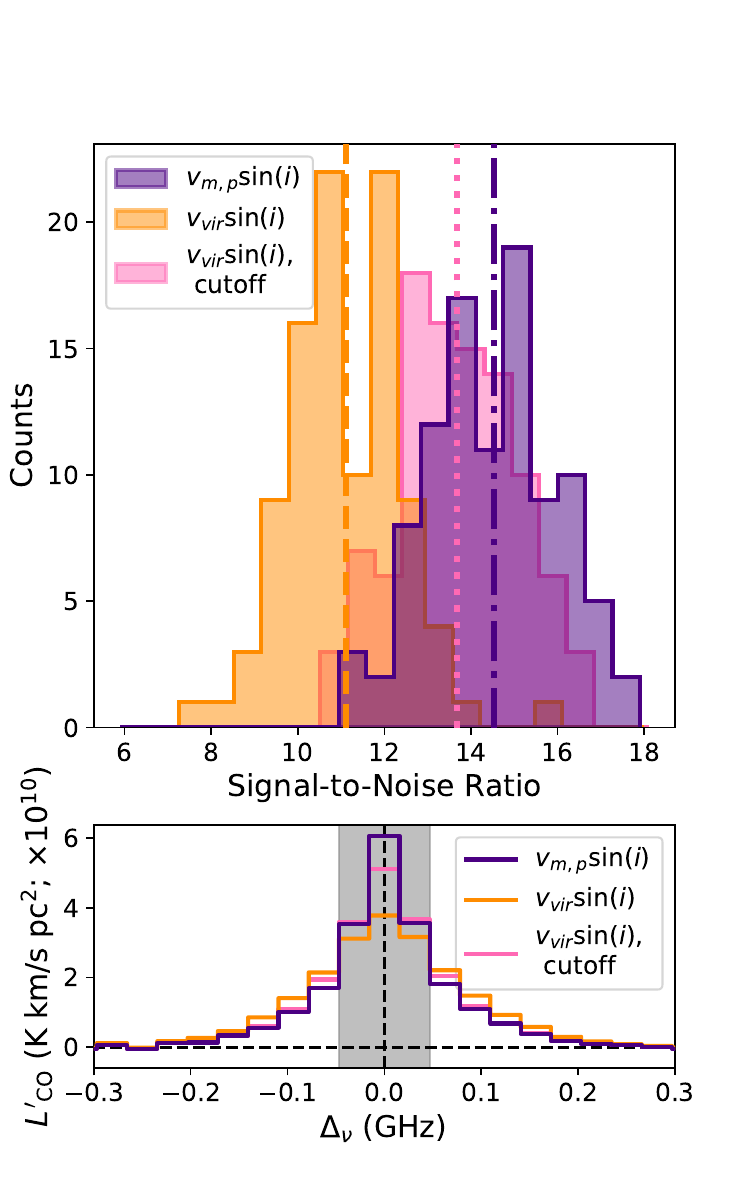}
    \caption{\textit{Top:} Histogram of the output S/N ratio for each prescription for astrophysical line broadening of CO. The different prescriptions are described in \S\ref{sec:simpipeline:simmap}. Mean values are indicated with vertical lines. \textit{Bottom:} Average stacked spectrum across simulation realisations for each prescription. The central 93.75 MHz, which is integrated across to determine the final stack luminosity, is shown in shaded grey. }
    \label{fig:line_broadening_snr_hists}
\end{figure}

Because CO linewidths are not well characterized at $z = 3$, we use the three theoretical prescriptions described in \S\ref{sec:simpipeline:simmap} to test the effects of line broadening on the stack. For each prescription, we perform stacks on 99 different simulation realisations. Unlike in the redshift uncertainty case, we do not attempt to correct for the CO linewidth prescription when determining the spectral aperture width of the stack -- we use three spectral channels, or $91.75\ \mathrm{MHz}$ in all cases. This is because we do not know a priori how the CO linewidths will behave at $z\sim 3$, (whereas the redshift uncertainty of a galaxy catalogue can typically be measured). It is possible that this effect could be measured using more advanced signal extraction techniques (than the straight sum we are currently using) in the spectral axis. The output S/N ratios for the different simulation realisations are compared in Figure \ref{fig:line_broadening_snr_hists}. Also shown are the average stacked spectra (across simulation realisations) for each line broadening prescription.

We find that using the virial velocity of the DM halo as the rotational velocity predicts the worst S/N ratio for the stack, where either using \vmp\ or the virial velocity with a cut-off predict higher S/N ratios. The difference in S/N ratio is due to the stacked linewidth being smaller for the \vmp\ method than for either of the \vvir-based methods, as can be seen in their average stacked spectra. The average FWHM of the stacked spectral line is 596 km/s using the \vmp\ method. The \vvir\ method without a 1000 km/s cut-off yields a line FWHM of 1212 km/s, and adding a cut-off reduces the FWHM to 813 km/s. 

This is consistent with the virial velocity method assigning high-mass DM halos much higher linewidths than have been observed (see Appendix \ref{app:line_broadening}). These high-mass halos are also likely the brightest halos and thus contribute more strongly to the stack. If the high-mass halos have larger linewidths, the stack as a whole is likely to be broader and thus have more signal falling outside of the spectral aperture used to calculate the final S/N ratio. Both the cut-off virial velocity and the \vmp\ prescription allow for narrower spectral lines at high halo masses.

\subsection{Interloper contamination}\label{sec:simresults:interlopers}

While we have thus far been assuming a perfect LIM map and galaxy catalogue, with only uncorrelated noise being added to the map, this is unlikely to be the case in a real intensity mapping situation. Potential sources of contamination to the stack include foreground or background interloper emission being present in the LIM map (for example, the COMAP data cubes will contain CO(2-1) emission from galaxies at $6<z<8$, which is expected to be roughly an order of magnitude fainter than the target CO(1-0) signal; \citealt{breysse2021_comapEOR}, \citealt{chung2024_globalsignals}), and interloper galaxies being present in the galaxy catalogue (due to either noise peaks or \textsc{[OII]}-emitting foreground galaxies). We explore both cases here. 

\subsubsection{LIM interlopers}\label{sec:simresults:interlopers:foregrounds}
\begin{figure}[ht!]
    \centering
    \includegraphics[width=0.95\linewidth]{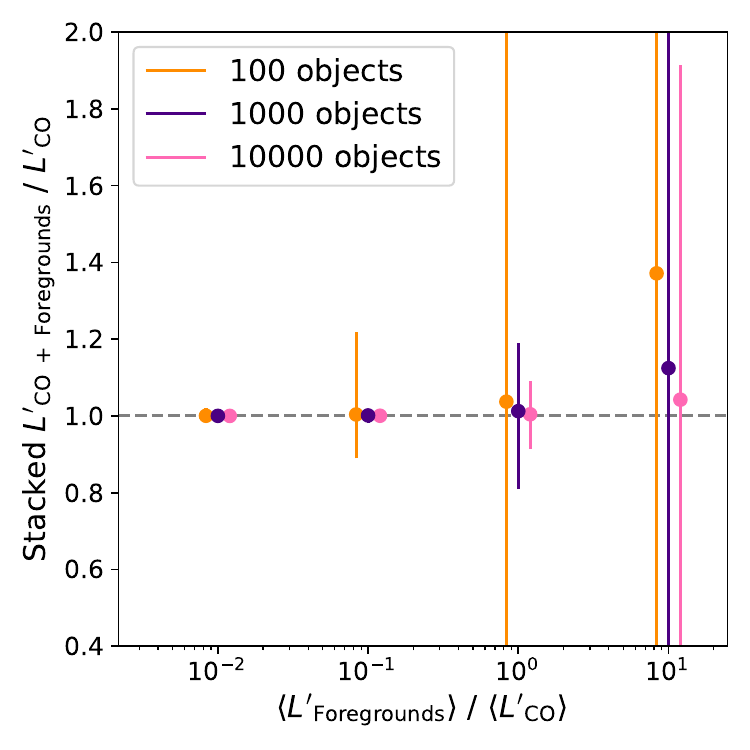}
    \caption{Average ratio of stack luminosity with added spectral line interlopers to the stack luminosity without interlopers for various scalings of the interloper emission. For each scaling, a CDF was created for the ratios from different simulation realisations, and the 95\% confidence interval is shown.}
    \label{fig:foregrounds}
\end{figure}

In order to test the robustness of the stacking pipeline against background and foreground interloper line emission being included in the LIM maps, we add in simulated background line emission following the methodology outlined in \S\ref{sec:simpipeline:simmap}. For each of nine different LIM map realisations, we generate 11 different interloper iterations. We then perform stacks on the maps with interloper emission present, at 1000\%, 100\%, 10\%, and 1\% of the target emission, and compare the output luminosity to the luminosity without interlopers. We repeat this analysis for stacks with 100 and 10 000 objects, to see if a larger galaxy catalogue helps to mitigate the effects of spectral line contamination. The results are shown in Figure \ref{fig:foregrounds}.

We find that the stack should be robust against the presence of faint spectral line interloper emission in the LIM map, although it is more affected by extremely bright contamination. Interloper emission has two effects on the maps: it biases the stacked luminosity towards higher values (as it is functionally adding more line signal into the maps, although this signal is not correlated with the actual population of galaxies the stack is aiming to probe), and it introduces scatter, as the catalogue objects will randomly be associated with regions of stronger or fainter interloper emission. On average, the maps with interlopers output the same values as the maps without: at the level of contamination expected in COMAP (roughly 10\%), the average deviation from the uncontaminated stack values is less than 0.5\% for all catalogue sizes. Even in the most extreme case, where the strength of the interlopers is $10\times$ that of the target line, the average deviation is 37\% in the 100-object stack, but only 4\% in the 10 000-object stack, showing that larger galaxy catalogues are more robust against interloper emission.

The amount of scatter added to the maps increases as the interlopers become brighter. At the expected COMAP interloper strength, the (95\%) scatter is 20\% (2\%, 0.9\%) for the stack with 100 (1000, 10 000) objects. This increases to 120\% (19\%, 9\%) for the same stacks when the interloper is of equal brightness to that of the target line, and more than 100\% for the stacks on all catalogues except the 10 000-object catalogue when the interloper is $10\times$ brighter than the target emission. The scatter is not solely to higher luminosity values because the maps are mean-subtracted -- adding extra emission to the maps will bring the overall mean up, so regions with few interloper galaxies will have their CO luminosity brought down by the mean subtraction. If the stacked catalogue is more associated with voids in the interloper map, the stacked luminosity will be brought down. We note that this analysis involves no attempt whatsoever to mitigate interloper contamination -- using a map-based strategy to handle contamination in LIM maps \citep[e.g.][]{sun2018_timeforegroundremoval, karoumpis2024_FYSTforegroundmasking} would likely reduce this scatter.

\subsubsection{Catalogue false positives}\label{sec:simresults:interlopers:falsepositives}
As with the intensity map, galaxy catalogues (especially those based on a single spectral line) may include some objects that are not galaxies at the target redshift emitting the target spectral line, but come from other spectral lines being emitted at other redshifts, or more rarely noise peaks being picked up by the signal-detection algorithm. HETDEX, for example, is expected to contain some population of $z < 0.5$ $[\mathrm{O\ II}]$ emitters misidentified as LAEs. These false positive detections will serve to bring down the overall signal in the stack, as each false positive object will add only noise to the stack. From this, one would expect the signal in a stack with a fraction $f_\mathrm{FP}$ of false positive objects to be attenuated by $f_\mathrm{FP}$.

\begin{figure}[t]
    \centering
    \includegraphics[width=0.95\linewidth]{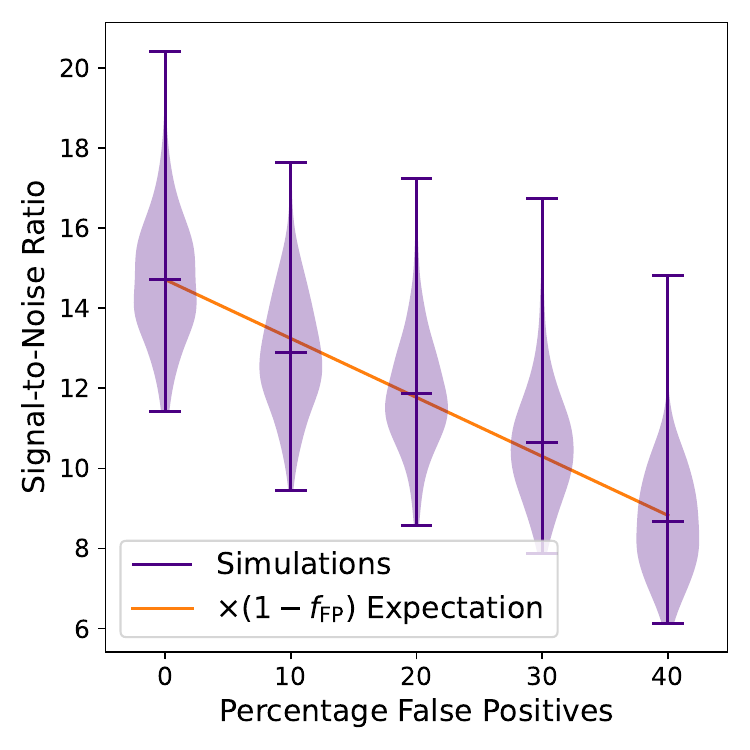}
    \caption{Signal-to-noise ratio of the stack as a function of the percentage of `false positive' interlopers included in the galaxy catalogue. Plotted in orange is the theoretical expectation, assuming the value of the map is identically zero at the (random) location of each false positive.}
    \label{fig:false_positives}
\end{figure}

In order to test this prediction, we stacked on catalogues where 0, 10, 20, 30, and 40\% of the catalogue was replaced with false positive (random) values (see \S\ref{sec:simpipeline:simgalaxycatalogue}). The results of these stacks are shown in Figure \ref{fig:false_positives}. We find that the attenuation of the stack due to the inclusion of false positive detections in the catalogue agrees well with the theoretical expectation, with some scatter. The scatter is slightly higher above the theoretical expectation, suggesting that this scatter may come from the random `false positive' catalogue entries falling onto brighter regions (regions of higher signal) of the map.

\section{Discussion}\label{sec:discussion}

\subsection{Ideal experimental setup}\label{sec:discussion:experimentalsetup}

\begin{table}[]
\caption{
    \label{tab:best_params} Summary of optimal parameters for the stack S/N ratio.}
    \centering
    \begin{tabular}{cc}
    \hline
   Parameter & Best value\\
   \hline 
   \multicolumn{2}{c}{\textit{Stack parameters}} \\
   \hline
       $N_\mathrm{spax}$  & 7\\
       $N_\mathrm{chan}$ & 3  \\
       $N_\mathrm{obj}$ & As high as possible \\
       $\Delta_z$ & 0 km/s  \\
   \hline 
   \multicolumn{2}{c}{\textit{LIM experimental parameters}} \\
   \hline
        $\delta_\nu$ & Sampling stack spatial extent \\
        $\delta_x$ & Sampling stack spectral extent \\
        $\theta_\mathrm{beam}$ & As small as possible \\
   \hline 
   \multicolumn{2}{c}{\textit{Astrophysical factors}} \\
   \hline
        CO model & L16 or C22 \\
        Ly$\alpha$ model & bright \\
        $\rho$ & 1.0 \\
        Line broadening & $v_\mathrm{m,peak}$ \\
   \hline 
   \multicolumn{2}{c}{\textit{Astrophysical factors}} \\
   \hline 
        Interloper emission & Tolerable to 10\% \\
        False positives & Tolerable  \\
    \hline 
    \end{tabular}
\end{table}

We list the results of the various tests of the LIM and galaxy catalogue experimental parameters in Table \ref{tab:best_params}. We find that the main consideration when selecting a galaxy catalogue for stacking is that the catalogue should have as many objects as possible. This drives down the noise in the stack by $\sqrt{N_\mathrm{obj}}$, but also helps to protect the stack against interloper emission in the LIM map. The uncertainty in the catalogue redshifts is also important, as any uncertainty spreads the signal over more noise and reduces the stack S/N ratio. Redshift uncertainties up to $\Delta_\mathrm{v}\sim 1000$ \kms\ may be tractable for smaller catalogues. High-precision photometric catalogues may also be useable, provided the large catalogue redshift uncertainty is offset by a significant increase in the number of catalogue objects across the LIM experiment's footprint.

The masses of the DM halos selected by the galaxy catalogue are also very important -- a galaxy catalogue that selects high-mass halos can improve the S/N ratio of the stack (when compared to a less-biased, Ly$\alpha$-like catalogue) more than any other single factor. We also note that galaxy surveys targeting different galaxy types could also be combined to maximize the collective $N_\mathrm{obj}$. The expected output signal would then become an average of the expected signal from each survey tracer, weighted by the number of objects included in each. In this way it may be possible to combine the benefits of, for example, a LAE survey with high $N_\mathrm{obj}$ and a more biased quasar survey with fewer objects. This may be especially valuable in the current LIM regime, where experiments are primarily searching for a detection, and thus the S/N is the primary concern.

Similarly to increasing the number of objects in the catalogue, the simplest way to adjust the LIM data to improve the sensitivity of the stack is to decrease the RMS noise in the data, through increased integration time or decreased $T_\mathrm{sys}$. This is not something we pursue in this work. Instead we explore the resolution and pixelisation of the map, given a fixed integration time. We find that the stack becomes less viable when the channel width of the LIM data is wider than the FWHM of the stacked spectral line -- LIM experiments with poorer spectral resolution may find stacking more difficult. Marginal improvements are also possible through having smaller pixel sizes in each axis of the LIM data, primarily because the stacked objects are able to be lined up more precisely when the central pixel in the stack is smaller.

As an additional point of consideration, most LIM experiments are designed to optimize for the auto-correlation power spectrum. None of the stack-driven experimental preferences that we list above will negatively impact power spectrum analysis, particularly at the high-$k$ end also probed by the stack. Improving the resolution, both spatially and spectrally, will increase the number of accessible $k$-modes without changing the noise power spectrum (the increased noise per voxel will be offset by the increased number of voxels, \citealt{li2016_comodelling}), thus improving the overall power spectrum signal-to-noise. In both cases, the primary constraints preventing this improved resolution are instrumental.

Finally, we explore the effects of various $L(M_\mathrm{h})$ models for the LIM tracer spectral line. While we framed this in the context of the largely unconstrained CO model space, this could equally be used as an exploration of which LIM tracers are most ideal for a stacking analysis. While a brighter tracer will automatically improve the stack luminosity -- any bulk increase to the tracer luminosity will result in a commensurate increase in the stack luminosity -- we also test models that have varying luminosity distributions across mass. We find that a tracer that is bright in the high-mass end (e.g.~C22) or a tracer that is bright in the low-mass end (e.g.~L16) each give greater S/N ratios than one that is a more moderate luminosity across all masses.  If selecting a LIM tracer specifically for a stacking analysis, this should be the primary consideration -- we find that the correlation on an individual halo level between the LIM tracer luminosity and the galaxy catalogue luminosity is not nearly as important. In order to explain these two effects, we must look in more detail at the halos contributing to the stack.

\subsection{Origin of the stack luminosity originate}\label{sec:discussion:clustering}

\begin{figure}
    \centering
    \includegraphics[width=0.98\linewidth]{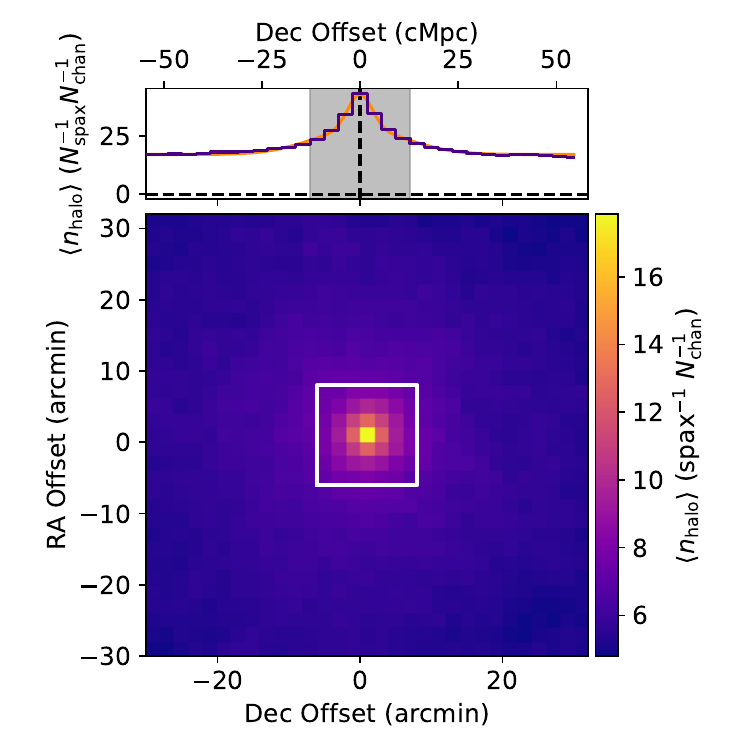}
    \caption{\textit{Bottom:} Number of DM halos included in each spaxel of the 2D spatial profile of the stack (i.e.~summed across the $N_\mathrm{chan}=3$ frequency channels of the stack aperture), averaged across cutouts. The actual catalogued DM halo is included in this count, but is outnumbered by more than an order of magnitude in the central spaxel alone. There are a significant number of DM halos in spaxels extending beyond the central spatial aperture. \textit{Top:} Spatial profile generated by integrating the $\langle n_\mathrm{halo}\rangle$ map across the three channels making up the stack aperture in the RA direction. Unlike the regular simulations, this is not mean-subtracted, and thus the baseline number of halos is high. A double-Gaussian fit to this profile is shown in orange.}
    \label{fig:n_halo_histogram}
\end{figure}

In order to understand the behaviour of the stack with parameters such as the CO $L(M_\mathrm{h})$ function or the correlation between the two tracers' scatter in luminosity, we explore the underlying distribution of halos, and how they drive the returned stack luminosity. Firstly, we explore where the DM halos that are contributing to the stack are located. We do this by generating a 3D histogram of the positions of all DM halos (without performing any cuts on $M_\mathrm{h}$, $L_\mathrm{CO}$, or $L_\mathrm{Ly\alpha}$, or weighting by any of the above) in a simulated stack run performed using default parameters. We then cut out the regions surrounding the DM halos included in the mock galaxy catalogue. We sum over the central three frequency channels (as is done when generating the actual 2D images and spatial profiles), yielding the DM halo density per spaxel, $n_\mathrm{halo}$, in the 2D image. We then average these halo density profiles across the cutouts associated with all 1000 catalogue objects included in the stack. The resulting distribution is shown in Figure \ref{fig:n_halo_histogram}. 

We find that on average 217 halos are contributing to the measured $\langle L_\mathrm{LIM} \rangle$ across the stack aperture. In the central voxel alone, there are, on average, ten halos contributing. There is also a fairly high baseline number of halos, with $\sim 20$ halos present even at the outskirts of the spatial profiles. This is consistent with the fact that the spatial profiles show luminosity extending nearly the full width of the stacked cubelet, even in the mean-subtracted simulated maps (we note that the halo number counts in Figure \ref{fig:n_halo_histogram} are not mean-subtracted in any way, unlike the actual mock data cubes). At least with the COMAP experimental parameters, we are operating on a broad enough resolution that DM halos can almost never be resolved individually.

We also extract a 1D spatial profile from the $\langle n_\mathrm{halo} \rangle$ map, and fit this profile with a two-Gaussian fit as is done for the actual simulated data in Figure \ref{fig:spatial_profile_fits}.  The narrow and broad components of the profile fit have standard deviations of $1.7'\pm0.1'$ and $6.6'\pm0.4'$ respectively. We note that the values of $3.2'$ and $9.6'$ reported for the full simulations are convolved with the $4.5'$ beam, where this $\langle n_\mathrm{halo} \rangle$ profile is not -- after deconvolving from the beam, the widths of the two Gaussian components in the full simulated stacks are $2.5'$ and $9.4'$. The broad component agrees within error between the two profiles, but the narrow component is narrower in the $\langle n_\mathrm{halo} \rangle$ stack profile than the full $L_\mathrm{LIM}$ version. This could be due to halos at radii slightly offset from the central objects being less frequent but brighter in CO.

\subsection{What the stack reveals about the galaxies contributing to the signal}\label{sec:discussion:comodels}
Though we have found that the stack signal originates from the large-scale clustering of galaxies, the extent of that clustering (and the centring of the stack objects in relation to it) will depend on the CO and Ly$\alpha$ luminosity properties of the galaxies being stacked. As we have found, the output stack luminosity depends heavily on the mass (and thus the bias) of the halos being targeted, and also depends on the CO luminosity at both the low- and high-mass ends of the DM mass function.

As such, we look into the luminosity contribution to the stack from both the catalogued objects themselves and their neighbours. We do this using the simulated catalogues of halo masses and CO luminosities, before generating any mock observations. We isolate two populations of halos that contribute to the stack. Firstly, we take the DM halos actually selected by the galaxy survey (via their \lya\ luminosity) to be stacked on. The CO luminosity of each of these halos corresponds to the first term in Equation \ref{eqn:stack_avg_luminosity}. Secondly, we took all of the halos that are not specifically selected by the galaxy survey but will still be counted in the stack due to their proximity to a selected halo. Practically, this amounts to all halos within 3.5 pixels along any axis of the LIM map. The contribution from these halos (the second term in Equation \ref{eqn:stack_avg_luminosity}) will be the sum over the $L'_\mathrm{CO}$ of all neighbours of a given selected halo.

\begin{figure*}
    \centering
    \includegraphics[width=0.98\linewidth]{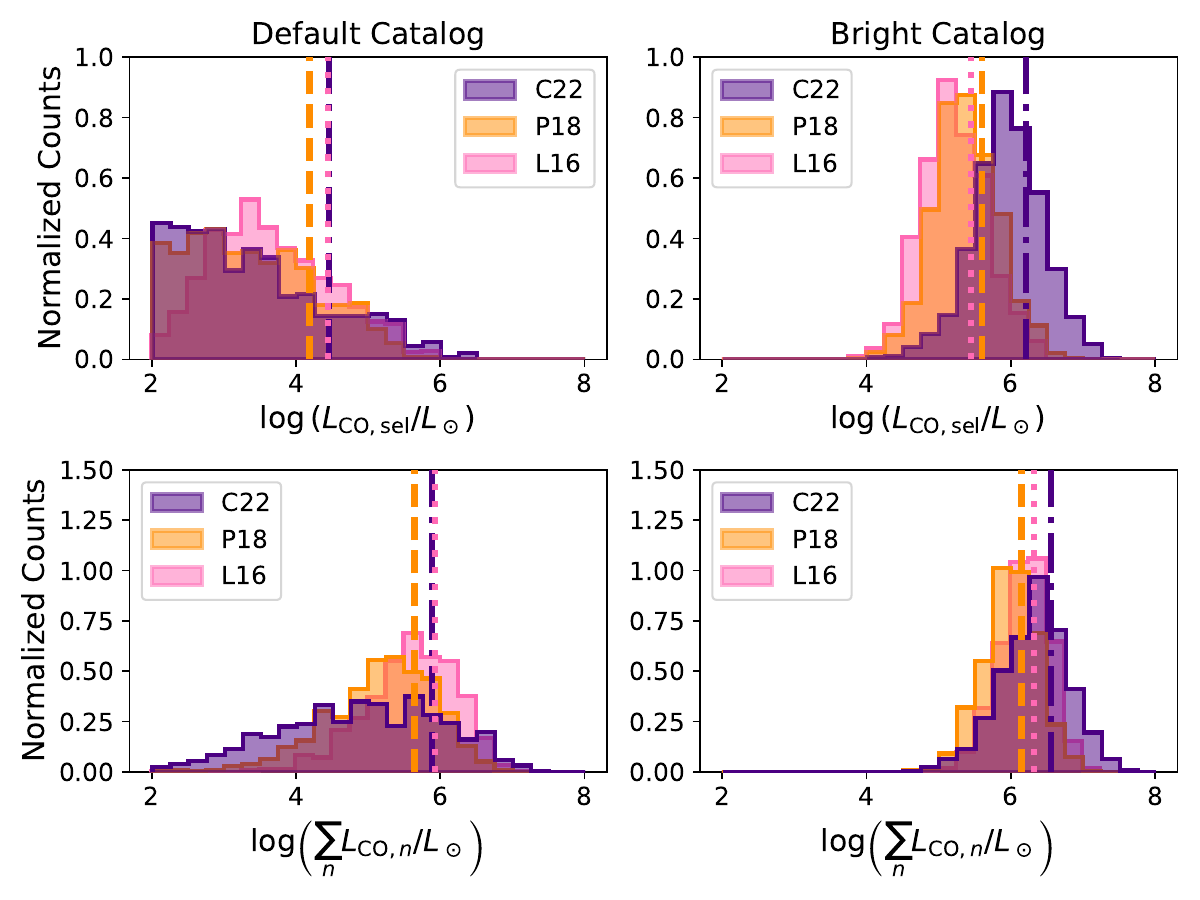}
    \caption{Comparison of the terms contributing to the overall stack luminosity for a single simulation realisation using the three different CO prescriptions. The top panel shows histograms of the $L_\mathrm{CO}$ of each DM halo included in the galaxy catalogue and thus directly selected to be included in the stack. The bottom panel shows the summed $L_\mathrm{CO}$ of all the halos neighbouring a given selected halo -- these halos also fall into the stack aperture and thus contribute to the stack as well. In both panels, the (linear) averages are shown as vertical lines.}
    \label{fig:default_stack_halo_breakdown}
\end{figure*}

For a single stack realisation, we plot a histogram of the $L_\mathrm{CO}$ values of each halo in the first population, and a histogram of the sum over the neighbouring halos of a given selected halo in Figure \ref{fig:default_stack_halo_breakdown}. We repeat this analysis for each of the CO luminosity prescriptions. We also plot the (linear) averages over all apertures included in the stack as vertical lines, keeping to the default Ly$\alpha$ catalogue model. The first, and most important, thing that we note from this analysis is that the stack luminosity is almost entirely driven by the neighbours of the halo selected by the galaxy catalogue, rather than the selected halo itself. For the default catalogue model, the integrated luminosity of the neighbouring halos for a given cutout is on average nearly two orders of magnitude greater than the luminosity of the selected halo. This agrees with the finding that there are on average more than 200 other halos contributing to the stack luminosity (Figure \ref{fig:n_halo_histogram}), and explains the fact that the stack signal is more extended in all three dimensions than a point-source contribution (\S\ref{sec:simresults:stackparams}).

We find that for both the neighbouring halos and the selected halo itself, the value predicted for P18 is lower than either of the other two prescriptions. In the case of the first term, this seems to be because the C22 value is driven by contributions from extremely bright high-mass values, and the L16 value, though its high-mass halos are not as bright, has many brighter low-mass halos. The P18 model has neither particularly bright high-mass halos nor low-mass halos. This also seems to be true of the second term, which is primarily a sum over lower-mass halos: the L16 model is brighter on average ($\sum_n L_{\mathrm{CO},n}$ is peaked high, with narrow spread), and the C22 model has sufficient bright contributors to drive its luminosity up (its spread in $\sum_n L_{\mathrm{CO}.n}$ is much larger, but it contains the brightest outliers). 

As a comparison point, we repeat this exercise for the bright, AGN-like galaxy catalogue, that selects more massive halos. This leads to two separate qualitative changes -- the selected halos are considerably more massive and thus brighter (the average luminosity increases by almost two orders of magnitude for each CO model), and they are more biased, so they have more (and brighter) neighbours. In this case, the contribution from neighbouring halos is still greater than the contribution from the catalogued halos themselves for all models, but the contributions are much more comparable. We do not show an equivalent test for the faint catalogue model here, as its behaviour is extremely similar to that of the default model.

The variation between the CO models in average luminosity of the selected halos follows from the fact that more massive halos are being selected. The C22 model is brightest at high halo mass, followed by P18 and then L16 (Figure \ref{fig:co_lum_funcs}). This is the same distribution seen in the selected halos. The average contribution from the neighbouring halos to the overall luminosity does not change as dramatically as the selected halo contribution from the default Ly$\alpha$ catalogue to the brighter one, but they are still more luminous for each CO model. Additionally, the distribution between neighbours of selected halos for a given model is much tighter. This is likely because more of the neighbouring halos are present due to actual clustering and not random chance. The C22 model also dominates considerably more in this case, probably because the more massive, more biased selected halos are more likely to have massive neighbours, which will be very bright in the C22 model. 

The origin of the stack signal also explains the behaviour of the stack with varying $\rho$, seen in Figure \ref{fig:rho_luminosity_changes} (\S\ref{sec:simresults:astrophysics:rho}). Variations in $\rho$ do not strongly impact the output stack luminosity, whereas variations in the catalogue luminosity function do, suggesting $S(L_\mathrm{Ly\alpha}(M_{h},\rho))$ is mostly driven by $L_\mathrm{Ly\alpha}$ rather than $\rho$. This follows from the fact that the neighbouring halos in the stack have a greater contribution to the total stack luminosity than the selected halos themselves for the default catalogue model -- a direct anti-correlation between the scatter in the CO and Ly$\alpha$ luminosities would mean the brightest Ly$\alpha$ halos have lower CO luminosities, so the halos directly selected by the galaxy catalogue would be fainter overall, but the neighbouring halos have a variety of luminosities in both tracers. It is possible that the small variations come from the selected halos being slightly less massive (and thus less biased) on average when $\rho$ is negative.

Additionally, we found that varying $\rho$ has proportionally more of an effect on the output stack luminosity for the more stochastic, top-heavy `bright' catalogue luminosity function than for the flatter default one. Figure \ref{fig:default_stack_halo_breakdown} shows that, when the bright catalogue luminosity function is used, the actual selected halos are contributing to the overall stack luminosity almost equally to the neighbouring halos. Correlating or anti-correlating the scatter will thus have a larger effect on the overall stack, as these selected halos are where the correlation matters.

\subsection{The stack in comparison to other summary statistics}

Using the survey parameters that we default to here (which were intended to simulate realistic small-scale CO and Ly$\alpha$ distributions while still achieving a solid stack detection with as few catalogue objects as possible), the mean stack detection S/N is 14.5. For the same parameters, the expected detection S/N ratio (i.e.~ignoring sample variance, enabling a direct comparison) of the cross-spectrum over all $k$ is 18.7. This is roughly 1.3$\times$ the stack S/N ratio. We note for completeness that the auto-spectrum S/N ratio is 53.4 across all $k$-modes, although the parameters we use here have far too few catalogue objects spread across too wide an area to allow for a fair comparison between the auto- and cross-spectrum analyses.

It is expected that the cross-spectrum be more significant than the stack -- the stack is probing the same signal as the smaller-scale modes ($k \lesssim 0.25$) of the cross-spectrum, but in real space rather than Fourier space. The cross-spectrum is sensitive to both these small-scale modes and the larger modes that the stack is not and thus has access to information that the stack does not. Additionally, this implies that any of the optimisations for stack signal we have discussed in this work (whether astrophysical or experimental) will also yield the strongest cross-correlation signal possible at small scales. Thus, in terms of raw statistics, one might expect to see a cross-spectrum detection before a stack detection. However, as we discuss below, practical analysis considerations mean the ordering may not be so clear-cut.

In addition to increased significance, there are analysis benefits to having information over multiple spatial scales -- many of the astrophysical and cosmological parameters that may be extricable from a cross-correlation \citep[e.g.][]{pullen2013_wmapquasars_limcrosscorr, breysse2019_agnfeedback} are extremely degenerate in the stack. More detailed comparisons between the stack and the cross-power spectrum are outside of the scope of this work, but we imagine it would be very difficult to extract more information than this overall amplitude from a stacking analysis.

While the cross-spectrum is likely a more valuable analysis tool in an idealised case, in practice LIM experiments are subject to many real-world complications that will likely be more difficult to disentangle from the cross-spectrum than the real-space stack. These include effects such as on-sky survey asymmetries in either the galaxy catalogue or LIM survey (HETDEX, for example, has an extremely complicated window function, \citealt{gebhardt2021_hetdexoverview}), systematic errors (ground pickup, standing waves, etc., all of which preferentially appear on the larger scales that the cross-spectrum is more sensitive to), and, particularly in the case of 21~cm experiments, continuum foregrounds. None of these complications are insurmountable in the cross-power spectrum, and there exists a wealth of literature already on addressing them \citep[e.g.][]{keenan2021_copssstack,lunde2024_COMAPS2_PaperI, meerklass2025_autocorr}. Still, they are likely easier to address in the stack. This has recently been shown in \cite{chen2025_meerklassstack}, where the authors were able to directly isolate and identify a specific source of systematic error in the MeerKLASS 21~cm data using their stacking analysis. In a field where we are still searching for first detections, the simplicity of the stack is a major advantage, and the stack and the cross-spectrum have the potential to work very well together. We will explore the complementarity of these two analyses in more detail in future work.

\section{Summary and conclusions}\label{sec:conclusions}
In this work, we have developed a simulation framework for multi-tracer LIM analyses and applied it to three-dimensional stacking of LIM data on the positions of objects from a resolved galaxy catalogue. We find that although stacking is conceptually simple, many degenerate parameters may affect both the stack signal and its noise. These parameters could come from the galaxy catalogue or the LIM data itself and could be experimental or astrophysical in origin. We posed three framing questions in the introduction to this work, and we aim to answer them here.

\begin{enumerate}
    \item What experimental design is optimal for a stacking analysis? We find that the optimal galaxy catalogue for stacking is a spectroscopic one that (a) has a maximum number of objects included and (b) targets high-mass DM halos. The ideal LIM experiment has a spectral resolution that is as high as possible (it must at least sample the stacked linewidth) and uses a spectral line tracer that is bright across the entire halo mass range, particularly at high halo masses. Other factors such as the spatial resolution of the LIM experiment or the correlation between the two spectral line tracers on an individual halo level may also affect the stack S/N ratio, although only marginally. The stack is quite resilient to faint ($\sim 10\%$) interloper emission in the LIM data.
    \item Where does the signal in a stacking analysis originate? For all choices of the galaxy catalogue luminosity function, the majority of the signal in the stack is not from the galaxies actually included in the catalogue itself. Instead, the signal is dominated by the integrated luminosity of the DM halos neighbouring the catalogued galaxies. This is true by more than an order of magnitude for our `default' galaxy catalogue luminosity function and by a factor of a few for a `bright' AGN-like luminosity function. This makes the stack very useful as a tool for LIM, where the goal is to measure galaxy properties and cosmology via aggregated galaxy emission.
    \item What can the stack tell us about the properties of the galaxies contributing to the signal? A galaxy catalogue targeting more massive biased halos leads to an increase in the stack signal. The stack signal is also improved by increased correlation between the luminosities of the two tracers being used (on a single-halo level) and brighter CO emission on either the high- or low-mass end of the $L(M)$ function. However, all of these parameters are degenerate, and all other parameters would need to be known in order to constrain any single parameter.
\end{enumerate}

As much of the field is still in a detection phase, we have framed this work only in terms of maximising the stack S/N ratio. We believe that the stack will be a useful tool for detection of the LIM signal, as the stack is easy to compute and has all the benefits of a joint analysis. It will be particularly valuable as an accompaniment to an auto-correlation detection based on the same LIM dataset, validating the origin of this signal in high-redshift galaxies. The stack will also be very useful in concert with a cross-correlation of LIM data with galaxy catalogues. Once a detection is made, characterising each of the astrophysical factors explored here will become a major goal of LIM surveys. The output of the stack is primarily a single data point, and we have shown that many of these astrophysical factors will have degenerate effects on that data point. Cross-correlations are sensitive to more spatial scales than the stack and thus allow for a more detailed comparison with models. However, cross-correlation analyses are more sensitive to systematic errors and may in turn rely on stacking analyses to constrain those errors. We will explore the interplay between these other analyses in future work. Ultimately, stacking is simple and robust, and it has the potential to be a useful tool in the LIM toolbox.

\begin{acknowledgements}

This material is based upon work supported by the National Science Foundation under Grant Nos.\ 1517108, 1517288, 1517598, 1518282, 1910999, and 2206834, as well as by the Keck Institute for Space Studies under `The First Billion Years: A Technical Development Program for Spectral Line Observations'. HP's research is supported by the Swiss National Science Foundation via Ambizione Grant PZ00P2\_179934. JK was supported by the National Research Foundation of Korea (NRF) grant funded by the Korea government (MSIT) (RS-2024-00340759). 
JG acknowledges support from the Keck Institute for Space Science, NSF AST-1517108 and University of Miami.
\end{acknowledgements}

\bibliography{aa54780-25}{}
\bibliographystyle{aa}

\begin{appendix}

\section{Modelling of astrophysical line broadening}\label{app:line_broadening}

In Section \ref{sec:simpipeline:simmap}, we selected from several different theoretical models for CO line broadening at $z=3$ without much justification. Here, we attempt to provide a basis for our choice of the \vmp\ method from \cite{behroozi2019_universemachine} through comparison with observations of CO-emitting galaxies. Ideally this comparison would be made directly with observations of CO(1--0) at $z=3$, but these observations are so limited that we include other CO transitions and lower redshifts in order to collect a larger sample. This yields an extremely heterogeneous collection of galaxies. We will aim to explore, and correct for, the effects of redshift and $J$ in future works.

\begin{figure}
    \centering
    \includegraphics[width=0.95\linewidth]{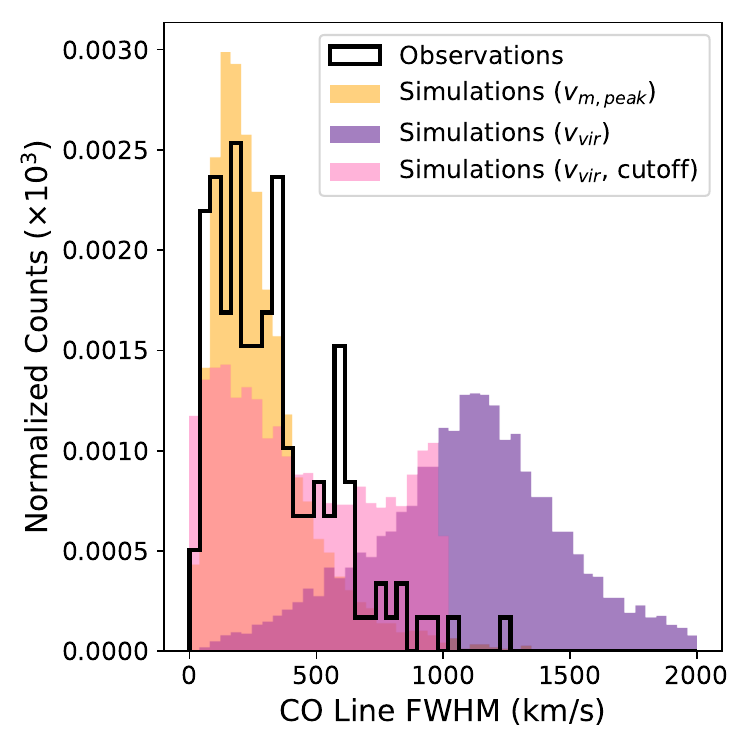}
    \caption{Relative occurrence of different CO line FWHM values across the simulated DM halo population for each of the line broadening prescriptions outlined in \S\ref{sec:simpipeline:simmap}. These are compared to observations (black line) of galaxies emitting in various CO lines up to $z=3.5$.}
    \label{fig:line_broadening_histograms_2}
\end{figure}

In total, we collected 119 high-z CO-detected galaxies from various sources:
\begin{itemize}
    \item ASPECS: We include 16 galaxies from the ALMA Spectroscopic Survey in the HUDF \citep[ASPECS,][]{aravena2019_aspecs,gonzalezlopez2019_aspecs,boogaard2019_aspecs,boogaard2020_aspecs}. This was a blind search for CO emission in ALMA's Band 3, so there is some confusion between CO transitions. The bulk of these objects are likely CO(2--1) emission from $z\sim1$ and CO(3--2) emission from $z\sim 2.5$. As we are only interested in the line FWHM for this application, we make no cuts on the confidence of the line identification. Line widths are reported directly as FWHM values in $\mathrm{km\ s^{-1}}$.
    \item VLAASPECS: This is an extension of the original ASPECS program using the Very Large Array \citep{riechers2020_vlaspecs}. We include the three confident CO(1--0) detections at $z\sim2.5$ from this survey in our sample. Again, linewidths are reported directly as FWHM values in \kms.
    \item PHIBSS: We include 53 galaxies from the IRAM Plateau de Bure high-z blue sequence CO(3-2) survey \citep[PHIBSS,][]{phibss2013_tacconi}. These are all CO(3--2) detections, at $z\sim1$, and are primarily massive, main-sequence star-forming galaxies. The linewidths for these galaxies have been converted to circular velocities ($v_\mathrm{c}$), so we undo this conversion to return the observed line FWHM values. This is done in two different ways for different galaxy types -- galaxies that have measured disc-like morphologies from HST and significant CO velocity gradients have been converted into $v_\mathrm{c}$ using the equation $v_\mathrm{c} = 1.3(\Delta v_\mathrm{blue-red}/2\sin i)$, and galaxies that have measured disc-like morphologies but no strong CO velocity gradient, or those with irregular morphologies, have been converted using $v_\mathrm{c} = \sqrt{3/8 \ln 2}\Delta v_\mathrm{FWHM}$. 
    \item PHIBSS2: We include an additional 43 galaxies from the second iteration of PHIBSS \citep[PHIBSS2,][]{freundlich2019_phibss2,lenkic2020_phibss2}, including both targeted and serendipitous detections. These galaxies include CO(1--0) emitters $0.5<z<0.8$, CO(2--1) emitters $z\sim1$, and some (16) higher-redshift, higher-$J$ CO emitters. The best-fit line FWHM is reported directly here.
    \item COLDz: We include the four CO(1--0) detections from the CO Luminosity Density at high-z \citep[COLDz,][]{pavesi2018_coldz,riechers2019_coldz} survey, all at $z\sim2.5$. The best-fit line FWHM is reported directly.
\end{itemize}
The FHWM for each of these galaxies is plotted in Figure \ref{fig:line_broadening_histograms_2}. They have an average linewidth of 385 \kms, with a long tail towards broader linewidths.

Also in Figure \ref{fig:line_broadening_histograms_2}, we plot the calculated line FWHM for each of the simulated galaxies included in the stack (i.e. all galaxies with $L_\mathrm{Ly\alpha}$ above the cut-off value for observation, $L_\mathrm{Ly\alpha,cut} = 8.26\times10^9\ L_\odot$) for each of the three line broadening prescriptions described in \S\ref{sec:simpipeline:simmap}. We note that these are not the only galaxies included in the simulated stack, as the neighbours of these galaxies that are near enough to fall into the stack aperture will also be averaged in. However, the selected halos are likely to be more massive than average and thus will have greater linewidths under any of the broadening prescriptions. They will thus dominate the stack linewidth, and are the relevant halos for this analysis.

We find that the \vmp\ prescription yields a distribution of line widths that matches the observations quite well. The \vvir\ prescription provides a distribution that peaks significantly higher than observations, and extends to much higher linewidths than are observed. Adding a 1000 \kms\ cut-off to the \vvir\ prescription does shift the average down to match more closely with observations, but the shape of the distribution is much more uniform, especially at high FWHM, than is actually observed.

\section{Model-dependence of the stack spatial extent}\label{app:spatial_extent}
As outlined in \S\ref{sec:simresults:stackparams:spatial} and \S\ref{sec:simresults:map:spatial_beam}, the extent of the signal in the stack cannot be explained solely by point sources at the location of the objects in the galaxy catalogue. This is particularly interesting in the angular axes, where the only broadening factors are beam smoothing and astrophysical clustering. If the large-scale signal visible in the stacks is indeed coming from clustering effects, the extent of this signal is dependent on both the CO and Ly$\alpha$ model chosen. While it is unlikely that the sensitivity of current and upcoming LIM experiments will allow for measurement of this spatial extent in detail in the near future (thus making a full exploration of its model-dependence outside the scope of this work), there will still be cosmological information in this distribution that may be exploitable in the longer-term.

\begin{table*}[!ht]
\caption{\label{tab:spatial_extents} Standard deviation values from double-Gaussian fits to spatial profiles of stacks generated using different model configurations.}
    \centering
    \begin{tabular}{cccccc}
    \hline
       CO model  & Ly$\alpha$ model & \multicolumn{2}{c}{$\sigma_\mathrm{narrow}$} & \multicolumn{2}{c}{$\sigma_\mathrm{broad}$} \\
            & & (arcmin) & (Mpc) & (arcmin) & (Mpc)\\
       \hline
       C22 & Default & $2.47\pm0.10$ & $4.5\pm0.2$ & $9.2\pm0.4$ & $16.7\pm0.8$\\
       P18 & Default & $2.59\pm0.07$ & $4.5\pm0.2$ & $10.2\pm0.4$ & $18.4\pm0.7$ \\
       L16 & Default & $2.53\pm0.07$ & $4.7\pm0.1$ & $9.9\pm0.4$ & $18.0\pm0.6$\\
       C22 & bright & $1.77\pm0.04$ & $3.23\pm0.07$ & $8.4\pm0.3$ & $15.3\pm0.6$\\
       C22 & faint & $2.44\pm0.07$ & $4.4\pm0.1$ & $11.5\pm0.5$ & $20.8\pm1.0$\\
       \hline
    \end{tabular}
    \tablefoot{The value $\sigma_\mathrm{narrow}$ always refers to the narrower component and $\sigma_\mathrm{broad}$ to the broader component.}
\end{table*}

\begin{figure}[t]
    \centering
    \includegraphics[width=0.98\linewidth]{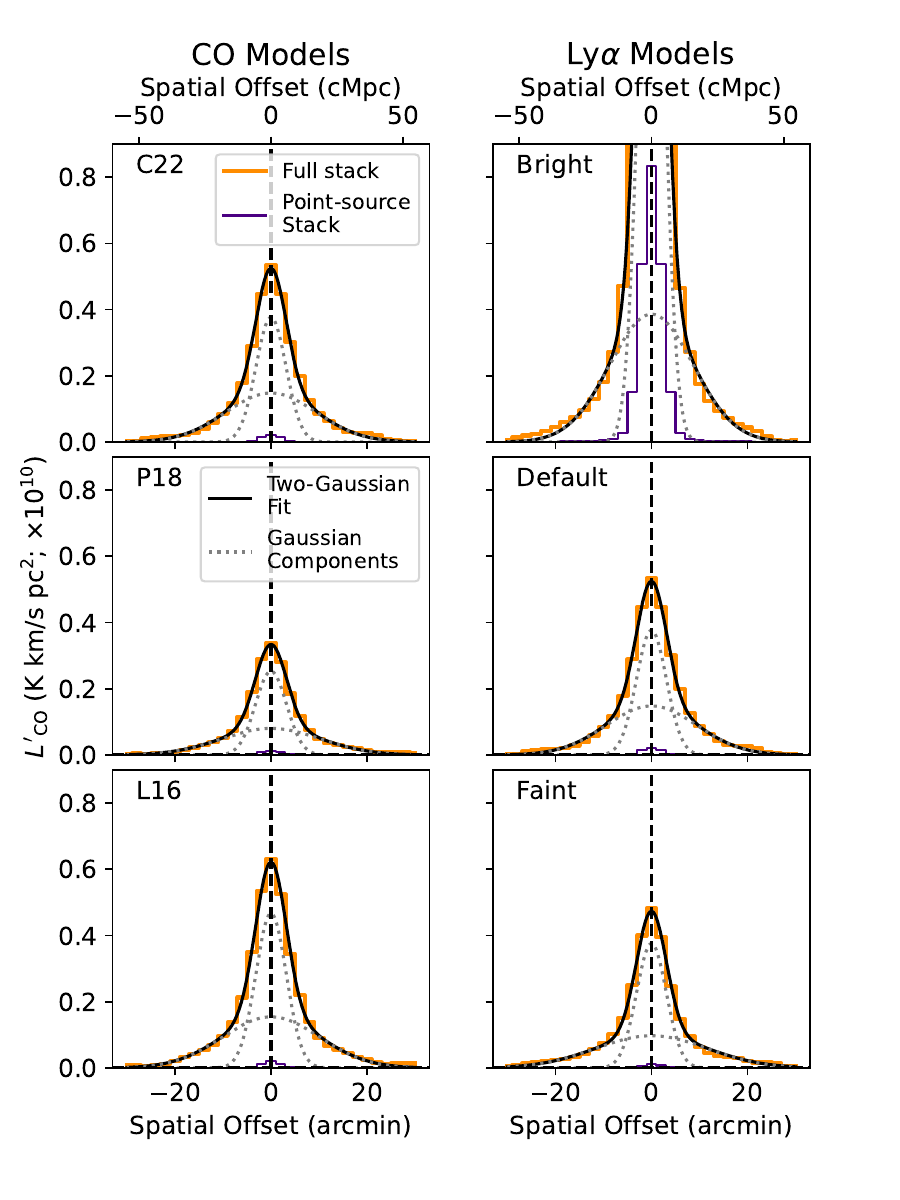}
    \caption{One-dimensional spatial profiles of stacks performed on simulations generated using various CO (left) and Ly$\alpha$ (right) models. For each profile, we plot the stack on the full simulated map (the orange lines) as well as a stack performed on a simulated map that contains only the DM halos included in the galaxy catalogue being stacked on (so there is no contribution at all from neighbouring halos; the indigo lines). We also plot a two-Gaussian fit (black lines) as well as the two Gaussian components individually (grey dotted lines).}
    \label{fig:co_lya_spatial_extents}
\end{figure}

Thus, we performed an initial characterisation by fitting two-Gaussian profiles to the spatial profiles of simulated stacks following \S\ref{sec:simresults:stackparams:spatial}.  For each stack exploring a different CO model, we adopt the default Ly$\alpha$ model. In those stacks testing different Ly$\alpha$ models, we adopt the fiducial C22 CO model. We also plot the average spatial profile generated from a set of nine catalogue-only simulation runs for each model. This is done following \S\ref{sec:simresults:stackparams} -- we set the $L_\mathrm{CO}$ of every halo not included in the galaxy catalogue to zero, to show the contribution from the catalogued halos themselves. The results of this are shown in Figure \ref{fig:co_lya_spatial_extents}. For most models, the contribution from the catalogued halos themselves is nearly negligible compared to the rest of the field.

The standard deviation values returned from the two-Gaussian fits are tabulated in Table \ref{tab:spatial_extents}. Unlike in \S\ref{sec:simresults:stackparams:spatial} above, we deconvolve out the $\mathrm{FWHM} = 4.5'$ beam from the standard deviation values before reporting them. We find that changing the CO model does not significantly change the spatial extent of the stack signal, while changing the Ly$\alpha$ model does result in some significant changes. This is because altering the CO model does not alter the DM halo population underlying the stack signal, as this is determined by which halos are included in the catalogue, which is set only by the Ly$\alpha$ luminosity (through $S(L_\mathrm{Ly\alpha}(M_{h,i},\rho))$, of the halos following Equation \ref{eqn:stack_avg_luminosity}). Thus changing the Ly$\alpha$ prescription used to generate the simulated catalogue alters the underlying population of DM halos behind the stack signal.

The standard deviation of the narrower Gaussian profile, $\sigma_\mathrm{narrow}$, corresponds to roughly $4.5\ \mathrm{Mpc}$ for most models. The only notable outlier is the AGN-like (`bright') Ly$\alpha$ model, which has a significantly smaller $\sigma_\mathrm{narrow}$. This is likely because the bright Ly$\alpha$ model has the most contribution from the actual catalogued objects (Figure \ref{fig:co_lya_spatial_extents}), so its narrow Gaussian component is the closest to a point-source fit. The broad component's standard deviation $\sigma_\mathrm{broad}$ is significantly offset from the default for both the bright and the narrow Ly$\alpha$ models. The bright model again has a narrower broad component than the default model, possibly because it is more heavily clustered. In contrast, the faint Ly$\alpha$ model has a much wider spatial extent than the default configuration. The stacks on catalogues generated using the faint model are centred around lower-mass halos than either of the other catalogue models' stacks. These halos are less biased, and thus stacks of their surroundings show less clustering than other models.

\end{appendix}
\end{document}